\documentclass[journal]{IEEEtran}
\usepackage{indentfirst}

\usepackage{diagbox}
\UseRawInputEncoding

\usepackage{cite}

\usepackage{multirow}

\usepackage{subfigure}
\usepackage{caption}

\ifCLASSINFOpdf
  \usepackage[pdftex]{graphicx}
  \graphicspath{{../pdf/}{../jpeg/}}
  \DeclareGraphicsExtensions{.pdf,.jpeg,.png}
\else
  \usepackage[dvips]{graphicx}
  \graphicspath{{../eps/}}
  \DeclareGraphicsExtensions{.eps}
\fi
\usepackage{amsmath}
\usepackage{algorithm}
\usepackage{algorithmic}

\usepackage{array}

\ifCLASSOPTIONcompsoc
\usepackage[caption=false,font=normalsize,labelfont=sf,textfont=sf]{subfig}
\else
\usepackage[caption=false,font=footnotesize]{subfig}
\usepackage{fixltx2e}
\usepackage{float}
\usepackage{stfloats}
\begin{document}
%
\title{Budgeted Influence Maximization via Boost Simulated Annealing in Social Networks}
%
%
%

\author{Jianshe~Wu,~\IEEEmembership{Senior Member, IEEE,}
       Junjun~Gao, Hongde~Zhu, Zulei ~Zhang
\thanks{J. Wu, J. Gao, H. Zhu,and Z. Zhang are with the Key Laboratory of Intelligent Perception and Image Understanding, Ministry of Education, International Research Center
for Intelligent Perception and Computation, Xidian University, Xi��an 710071,China (e-mail: jshwu@mail.xidian.edu.cn;  1348007434@qq.com; hdzhu\_xidian@sina.com; m18292823852@163.com).}}


%
%

\markboth{Journal of \LaTeX\ Class Files,~Vol.~10, No.~1, April~2021}%
{Shell \MakeLowercase{\textit{et al.}}: Bare Demo of IEEEtran.cls for IEEE Journals}
%



\maketitle

\begin{abstract}
Due to much closer to real application scenarios, the budgeted influence maximization (BIM) problem has attracted great attention among researchers. As a variant of the influence maximization (IM) problem, the BIM problem aims at mining several nodes with different costs as seeds with limited budget to  maximize the influence as possible. By first activating these
seed nodes and spreading influence under the given propagation model, the maximized spread of influence can be reached in the network.

Several approaches have been proposed for BIM. Most of them are modified versions of the greedy algorithm, which work well on the IM but seems inefficient for the BIM because huge time consuming is inevitable. Recently, some intelligence algorithms are proposed in order to reduce the running time, but analysis shows that they cannot fully utilize the relationships between nodes in networks, which will result in influence loss.

Inspired by this, we propose an efficient method based on boosted simulated annealing (SA) algorithm in this paper. Three heuristic strategies are proposed to improve the performance and speed up the proposed algorithm. Experimental results on both real world and synthetic networks demonstrate that the proposed boosted SA performs much better than existed algorithms on performance with almost equal or less running time.

\end{abstract}


%
\IEEEpeerreviewmaketitle

\section{\textbf{Introduction}}
\IEEEPARstart{W}{ith} the development of information technology, many well-known large-scale network cites, such as Wechat, Facebook, Twitter and Youtube, have been widely used in  information dissemination and influence diffusion. The current researches indicated that people can be more willing to accept recommendation from their families and close friends than from other channels such as Handbill and newspaper \cite{Chen2010Scalable,Yu2015Friend}. This so called word-of-mouth social phenomenon can be utilized in the information spread for people to achieve a wide range of influence. Meanwhile, compared with offline social relationships, the online social relationships have more frequent connections with each others and are easier to be computed, which means companies may get higher profits with less investment.

As an important question originated from viral marketing, the IM problem plays a critical role in the influence spread \cite{Chen2020Efficient,2018In}. The original problem scenario is that one company tends to promote their products to consumers, they decide to pay some money or product discount to some online public figures for making advertisements to their followers \cite{Feng2020Neig}. There is no doubt that the company wants to get the most influence spread with a given number of selected seeds. Inspired by this social demand, Domingos and Richardson first investigated the problem. They translate it into a network structure and treat it as an optimization problem \cite{Domingos2001Mining}. Kempe et.al  first pointed out that the IM is an NP-hard problem and can only get approximate result \cite{Tardos2003Maximizing}. They put forward a greedy algorithm and achieved ($1-1/e$) optimal spread.

However, when trying to use the IM model into practical scenarios, it is found that the IM model can not describe the characteristics of real social networks in many cases. As a matter of fact, as public figures, they have greater social influence than ordinary people, which means that they deserve to be paid more money to promote products or make advertisements. This condition is different from the setting in the IM problem, which assumed that all seed nodes have uniform cost. Usually the company has a given limited budget instead of a given number of seeds. Nguyen and Zheng investigated the BIM problem firstly and a greedy algorithm was proposed \cite{Hallier2013On}. 

In fact, variants of the IM problem have been investigated recently, such as the time-constrained IM \cite{Tong2020time}, targeted IM \cite{Cai2020Tar,2016Tar}, community-aware IM \cite{2017Most}, opinion maximization \cite{2019Opinion}, etc. In BIM, each seed node has different costs. Leskovec et.al put forward an improved greedy algorithm to deal with the BIM \cite{Leskovec2007Cost}. Although several strategies have been made to reduce the runtime, it is impossible to overcome the intrinsic flaw of enormous computation in a greedy algorithm. Without using a greedy strategy, Han et.al. come up with a combination simulated annealing (Combination SA) for the BIM \cite{Han2014Balanced}, which reduces the runtime dramatically. In fact, the combination SA provides a mechanism which gets balance between the accuracy and efficiency, but it ignores the relationships among nodes.

In this paper, we propose a seed selection method based on boosted simulated annealing (Boost SA) to deal with the BIM problem by skillfully using the topology of networks. Our main contributions are listed as follows:

\begin{enumerate}
  \item By using the information of the node's connections, the accuracy is improved and the candidate space of seed nodes is  dramatically reduced, which reduces the runtime.
  \item The voting mechanism is introduced to strengthen random substitution process based on the idea of ensemble learning, which improves the accuracy. 
  \item Adaptive interrupt mechanism is also used to further reduce the runtime. With fixed iteration number, the SA may waste time on useless computation. The adaptive interrupt mechanism can stop the iteration if there is no further improvement.
\end{enumerate}
\quad Experiments on three real world networks and synthetic networks show that the proposed Boost SA algorithm performs much better than the state-of-art Combination SA algorithm \cite{Han2014Balanced} with equal or less runtime.

 The rest of this paper is organized as follows: Section 2  is the related works and motivations; Section 3 is the problem formulation and typical diffusion models. Section 4 provides the detailed of the proposed Boost SA. Section 5 is the experimental results and  analysis. Section 6 is the conclusion.

\section{\textbf{Related Works and Motivations}}
\subsection{\textbf{Related Works}}
\textbf{Influence maximization (IM) problem.} Without considering the budget, the IM problem is formulated as how to find $k$ nodes (as seeds)  such that the expected number of propagated nodes by the $k$ seeds is the largest as possible. As one of the main problems originated from viral marketing, the IM has attracted great attention among researchers. The IM was first studied by Domingos and Richardson \cite{Domingos2001Mining}. Kempe et.al  proved that it is NP-hard and can only get approximate solution \cite{Tardos2003Maximizing}. They put forward a greedy algorithm under the independent cascade (IC) model and linear threshold (LT) model. This algorithm gets ($1-1/e$) optimal spread with expensive computation. Chen et.al  come up with two methods \cite{Chen2009Efficient}, named as NewGreedy and MixedGreedy,  to improve the greedy algorithm. Kundu and Pal proposed a deprecation based greedy strategy \cite{KUNDU2015107}, in this strategy, they estimate the performance of each node and mark the nodes to be deprecated. They also proved that if the influence function is monotonic and sub-modular, this strategy can provide a guaranteed seed set.

In addition to these methods based on a greedy algorithm, some simple heuristic approaches are proposed \cite{Tardos2003Maximizing,Brin2012Reprint,SAITO2016985},  including selecting the seed nodes by their degree, centrality, and many other indexes to measure the importance of the nodes. These strategies are easy to carry out but inevitably ignore the connections between nodes. In order to deal with the expensive Monte Carlo simulations in greedy methods, Jiang et.al  introduce an influence approximation index called the expected diffusion value (EDV) to compute the potential influence spread of each node \cite{Jiang2011Simulated}. Meanwhile, they artfully combine SA with the proposed EDV.

As an intelligent algorithm, SA was proposed by Metropolis et.al in 1953 \cite{Metropolis2004Equation}. The main idea of SA is to simulate the process of metal annealing. Ref\cite{Jiang2011Simulated} is the first work utilizing SA for the IM problem, which contains two layers of circulation, the outer circulation is controlled by temperature parameter $T(t)$ and the inner circulation is controlled by constant parameter $q$. The method operates as follows:

\begin{enumerate}
  \item From the initial seed set $A$, initial temperature $t_0$,  calculate the influence spread $\sigma(A)$ by EDV.
  \item Create a neighbor set $A^{'}$ of $A$ and calculate $\sigma(A^{'})$. If $\Delta E=\sigma(A^{'})-\sigma(A) \geq 0$, replace $A$ with $A^{'}$; else if $\Delta E=\sigma(A^{'})-\sigma(A) < 0$, replace $A$ with $A^{'}$ by probability $exp(\Delta E / T) $. The iteration will go on until the temperature comes to the given threshold.
\end{enumerate}

Swarm intelligence methods are also proposed to solve the IM problem. Gong et.al proposed a discrete particle swarm optimization method (DPSO) which combines with the extended EDV and get improvements in both influence spread and runtime \cite{Gong2016Influence}.

\textbf{Budgeted influence maximization (BIM) problem.} The BIM is formulated as how to select a number of seeds with given limited budget to maximize the influence spread. Obviously, the IM problem is a special case of the BIM when the costs to activate the nodes are equal, thus the algorithms developed for the BIM can be directly applied to the traditional IM problem.

Leskovec et.al.  first  build up a method called Cost-Effective lazy Forward (CELF) for the BIM, which uses the submodularity property to speed up the algorithm and it is much fast than a simple greedy algorithm \cite{Leskovec2007Cost}. Nguyen and Zheng  identify the linkage between the computation of marginal probabilities in Bayesian networks and the influence spread, they propose a method by estimating the influence spread via belief propagation on a underlying directed acyclic graph\cite{Hallier2013On}.

Although several strategies have been made to speed up the greedy strategy, it is not a good choice due to its intrinsic flaw of expensive computation. Instead of using a greedy strategy for evaluating the influence spread of a node or node set, Han et.al  provide two strategies, named Billboard strategy and Handbill strategy for evaluating and selecting seed node, then by balanced using the two strategies via simulated annealing, a new method (Combination SA) is proposed\cite{Han2014Balanced}.

As shown in Fig.1. the main processes of Combination SA are summarized as follows.

 \textbf{Step 1}. Dividing the nodes of the network into two subsets by their degrees: the set of the nodes with the top $20\%$ largest degree (denoted as $T$) and the set of the nodes with the bottom $80\%$ smallest degree (denoted as $H$).

\textbf{Step 2}. Constructing the candidate seed set (named billboard set and denoted as $S_b$) by the billboard strategy, which iteratively selects the most influential node from $T$ by an influence indicator until the budget $B$ is reached. The influence indicator evaluates a node's influence with only the node's  1-hop and 2-hop neighbors information (see Eq.\ref{eq5}).

\textbf{Step 3}. Constructing the candidate seed set (named handbill set and denoted as $S_h$) by the handbill strategy, which iteratively selects the most cost-effective node from $H$ by an indicator until the budget $B$ is reached. The cost-effective indicator considers not only the node's influence but also its cost with it's 1-hop and 2-hop neighbors information (see Eq.\ref{eq6}).

\textbf{Step 4}. Set $S_b$  as the initial seed set $S$, then try to replace each seed node in $S$ with nodes in $S_h$ that has equal costs. Acceptance or rejection of the replacement is decided by SA. If accepted the replacement, go to Step 5.

\textbf{Step 5}. Randomly replace a node in $S$ with nodes in $S_h$  with equal costs. Similarly,  acceptance or rejection of the replacement is judged by SA. This operation  execute $q$ times.

When all nodes in set $S$ have been processed by Step 4 and Step 5, the algorithm is finished. As a replacement of traditional greedy algorithm, SA is a good choice to find a solution for the BIM problem. In fact, the Combination SA  has achieved good results on the BIM problem.

 Yang et.al. formulate the BIM as a Multi-Objective optimization problem and utilize Discrete Particle Swarm Optimization algorithm (MODPSO) to deal with \cite{Yang2018Influence}, which is much fast than a greedy algorithm but is slower than the Combination SA.

\begin{figure}
\begin{minipage}{0.5\textwidth}
  \centerline{\includegraphics[width=2.5in]{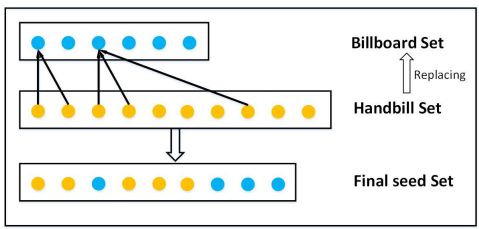}}
  \caption{Process of Combination SA Algorithm}
\end{minipage}
\end{figure}
\subsection{\textbf{Motivations}}

\textbf{Motivation 1:} Make use of the relations between nodes of $T$ and $H$.  As introduced above, Combination SA constructs two candidate sets: billboard set ($S_b$) and handbill set ($S_h$), which come from $T$ and $H$ respectively. The separation of $T$ and $H$ is merely based on the nodes' degrees. Thus, the separation of $S_b$ and $S_h$  ignores the relations (edges) between the nodes of $T$ and $H$, which results in unnecessary budget waste (performance descending). For example, a node is selected into $S_b$, its direct neighbor may be selected into $S_h$ if the edge information is ignored, which is a waste of budget. So, the first motivation is to construct only one candidate seed set.

\textbf{Motivation 2:} Reducing the searching space. As a whole, the two candidate sets including billboard set $S_b$ and handbill set $S_h$ are too large, iteratively search in which will result in long runtime. As introduced above, the billboard set $S_b$ is constructed by selecting the most influential node from $T$ one by one until the budget $B$ is reached. The handbill set $S_h$ is constructed by selecting the most cost-effective node from $H$ one by one until the budget $B$ is reached. The candidate set $S_b+S_h$ has nodes with total cost of $2B$, by constructing only one candidate set $C$ the search space can be reduced to nodes with total cost of about $1.5B$.

\textbf{Motivation 3:} The searching scheme of Combination SA is trying to replace each node in $S_b$ by several nodes in $S_h$, the resulted set is the obtained seed set. In fact, starting from different initial sets, after the replacing process by simulated annealing, the resulted seed sets are different. The third motivation is to start from $k$ different initial sets selected from the candidate set $C$ with beget $B$, obtains $k$ initial seed sets. By designing a voting strategy to obtain the initial seed set from the $k$  seed sets. The voting strategy is also unsed in obtain the final seed set, which improves the accuracy but increases the runtime.
\section{\textbf{Problem Description and Preliminaries }}
In this section, we first  formulate the BIM problem in mathematical language. Then typical diffusion models are introduced which describe the diffusion behavior in different scenarios. Finally, the influence indicators for a node or nodes set are analyzed. Some important notations are listed in Table \ref{table1} as follows:

\newcommand{\tabincell}[2]{
\begin{tabular}{@{}#1@{}}#2
\end{tabular}}
\begin{table}[H]
\centering
\caption{Notations}
\label{table1}
\begin{tabular}{|c|l|}

  \hline
  {Notation}&\multicolumn{1}{|c|}{Details}\\
  \hline
  $G(V,E)$ & \tabincell{l}{$G$ is the target network,$V$ and $E$ are the \\nodes and edges set in $G$ respectively.}\\
  \hline
  $S$ & Seed node set. \\
  \hline
  $B$ & The given constant budget. \\
  \hline
  $c(S)$ & The total cost of  seed set $S$. \\
  \hline
  $\sigma(S)$ & The influence spread of  seed set $S$. \\
  \hline
  $ce(v_i)$ & The cost-effective value of node $v_i$ \\
  \hline
  $N(v_i)$ & The set of direct neighbors of node $v_i$. \\
  \hline
  $N_{out}(v_i)$ & The set of direct out-degree neighbors of node $v_i$. \\
  \hline
  $N_{in}(v_i)$ & The set of direct in-degree neighbors of node $v_i$. \\
  \hline
\end{tabular}
\end{table}

\subsubsection*{\textbf{3.1	Formulation of the BIM Problem}}

Before formulating of the BIM problem, the ability of the influence spread of a single node and a seed set should be explained.

\indent \textbf{Influence spread.} The ability of influence spread of a single active node $v_i$ is defined as the expected number of nodes which can be activated by $v_i$, which is denoted as $\sigma(v_i)$. Similarly, the ability of influence spread of a seed set $S$ is defined as the expected number of nodes which may be activated by $S$, which is denoted as $\sigma(S)$.

\indent \textbf{Budgeted influence maximization (BIM).} A social network is usually modeled as a graph $G(V,E)$, where  $V=\{v_{1},v_{2},\ldots,v_{n}\}$ is the set of individuals (nodes) and  $E={e_{ij}}$ is the set of edges. For each $v_i\in V$, it has a cost $c(v_i)$ when it is selected to be a seed node.  Given the budget $B$, the BIM problem aims at obtaining the most influence spread $\sigma(S)$ by selecting a set of seed nodes $S$  under the budget constraint, which is  formulated as follows:
\begin{equation}
\label{bim}
\begin{cases}
S=\arg \max \sigma(S),\ S\in V, \\
$subject\ to$\ \sum_{v_i \in S} c(v_i)\leq B. \\
\end{cases}
\end{equation}

When $c(v_i) \equiv 1$, $\forall v_i \in V$, the BIM equals to the original IM problem, thus the BIM is a generalization of IM.

\subsubsection*{\textbf{3.2	Diffusion Model}}
The BIM problem can be seen as the combination of the seed selection and influence diffusion. There are several diffusion models to describe the influence diffusion process in previous literatures. Here we introduce three main typical diffusion models and some of their variants.

\textbf{Independent Cascade (IC) Model.} In this model, after a node (e.g. $v_{i}$) is activated in step $t$, $v_{i}$ has only one chance  to activate each of its inactive neighbors $(e.g.  v_{j})$ with probability in step $t+1$ \cite{Jiang2011Simulated,Han2014Balanced}. Whether $v_{j}$ is successfully activated by $v_{i}$ or not is independent of the history of diffusion before $v_{i}$ is activated. If node $v_{j}$ has $L$ ($L>=2$) neighbors which are activated in step $t$, $v_{j}$ will have $L$ chances to be activated by its neighbors sequentially in step $t+1$.

\textbf{Linear Threshold (LT) Model.} In this model, each inactive node $v_{j}$ is influenced by all its active neighbors according to the summation of edges�� weight: $p_{j}=\sum p_{ij}$, where $v_{i}$ is active. $\Theta_{j}$ is set as the threshold for activating $v_{j}$. If $p_{j} \geq \Theta_{j}$ in step $t$, node $v_j$ is activated in that step. The threshold $\Theta_{j}$ can be defined node-specific, e.g. uniformly distributed in the interval $[0, 1]$, or adopt an identical value like 1/2 \cite{Chen2010Scalable}. The process will stop until no more node can be activated.

\textbf{Triggering Model.} In this model, each node $v_{j}$ independently select a set of nodes from its neighbors according to some distribution, denoted as $T_{j}$. If $v_{i}$ is activated in step $t$ and $v_{i}\in T_{j}$, then $v_{j}$ will be activated in step $t+1$ \cite{Tardos2003Maximizing,Tang2015Influence}.

In recent years, there are researches about model independed for the IM problem \cite{Pan2020Multi} and model considering heterogeneous nodes \cite{2020An} . The same as that in Refs\cite{Jiang2011Simulated,Han2014Balanced},  in this paper, we study the BIM problem in the IC model. \\

\subsubsection*{\textbf{3.3	Influence Approximation Indicator}}
 Seed selection strategy is an important issue for the BIM problem. It's necessary to come up with an effective indicator to evaluate the influence of a node or nodes set. Since the computation of $\sigma(S)$ is NP-hard \cite{Zhang2010Estimate,Zhu2014Maximizing}, some simple indicators have been proposed.

   \textbf{Degree centrality} \cite{Tardos2003Maximizing}. The main idea is that if a node has more neighbors, it's more likely to have greater influence on the network. So degree centrality selects nodes with the maximum degree in descending order until the budget runs out. The degree of node $v_i$ is computed by:

      \begin{equation}
      d_{i}=\sum_{i\neq j} e_{ij}
      \end{equation}

      In Eq.2, if node $v_i$ and node $v_j$ have an edge, $e_{ij}=1$, otherwise $e_{ij}=0$.
Degree centrality may be the simplest indicator.

   \textbf{Expected diffusion value (EDV) for seeds set} \cite{Jiang2011Simulated}. EDV considers not only the degree of a node but also its activation probability, which is computed by:

      \begin{equation}
      \label{eq3}
      EDV(S)=k+\sum_{i\in N(S)^{}\setminus S} 1-(1-p)^{\tau(i)},
      \end{equation}

      In Eq.3, $k$ is the number of nodes in set $S$, $N(S)^{}$ is the set of direct neighbors of nodes in $S$, $v_i\in N{(S)}^{}\setminus S$ indicates that $v_i$ is a direct neighbor of nodes in $S$ but $v_i$ itself doesn't belong to $S$, $p$ is the activation probability with preset value, and $\tau(v_i)$ is the number of edges between $v_i$ and the nodes in $S$.

      In fact, Eq.\ref{eq3} only considers 1-hop nodes, in other words, the seeds and the direct neighbors of the seeds set.

  \textbf {2-hop influence spread for seeds set} $\sigma_2(S)$ \cite{Lee2014A}. As well known, the computation of $\sigma(S)$ is expensive, degree centrality ignores more long-distance information. A compromise method is only computing up to the 2-hop nodes, which results in a fast influence indicator $\sigma_2(S)$. Specifically, $\sigma_2(S)$  is  the nodes that may be activated by any a node in $S$ along a path equal or less than 2 hops. The indicator is computed by:
     \begin{align}
     \label{eq4}
      \sigma_2(S)&=\sum_{v_i\in S} \sigma_2(v_i) \nonumber \\
      &-(\sum_{v_i\in S} \sum_{v_l\in N{(v_i)}\bigcap S} p(v_i,v_l)(\sigma_{1}{(v_l)}-p(v_l,v_i)) )\\
      &-\chi, \nonumber
      \end{align}
where  $\sigma_2(v_i)$ is the nodes that may be activated within 2 hops by $v_i \in S$. Obviously, if a node is in 2-hop of $v_i \in S$ and is also in 2-hop of $v_j \in S$ ($v_i \neq v_j$), the node maybe counting two times in $\sum_{v_i\in S} \sigma_2(v_i)$, the seconde and the third terms aim at removing those repetitive nodes.

       The second term is the repetitive nodes in 1-hop. In details, $N{(v_i)}$ is the direct neighbors of  $v_i$, $p(v_i,v_l)$ is the probability that active node $v_i$ may activate inactive node $v_l$, $\sigma_1{(v_l)}=1+\sum_{v_k\in N{(v_l)}} p(v_l,v_k)$, which is 1-hop influence of node $v_l$.

       The third term $\chi$ is the repetitive nodes in 2-hop, which is computed by:
      \begin{equation}\nonumber
      \chi =\sum_{v_i\in S} \sum_{v_l\in N{(v_s)}\bigcap S} \sum_{v_d\in N{(v_l)}\bigcap S\setminus {v_s}} p(v_s,v_l)p(v_l,v_d),
       \end{equation}
       which represents repeated influence calculation of two seed nodes with a distance of 2-hops.

\textbf {Expected diffusion value (EDV) for a node}\cite{Han2014Balanced}.
 Han et.al. extend Eq.\ref{eq3} for evaluating the influence of a node by considering 2-hop nodes, which is computed by:

     \begin{align}
     \label{eq5}
      \sigma_2(v_i)&=1+\sum_{v_j\in N(v_i)} (1+outd(v_j)*p)*p \\\nonumber &-  \sum_{ v_{u},v_{k}\in N(v_i),e_{uk}\in E} p^3,
   \end{align}
where,  $outd(v_i)$ is the out-degree of $v_i$. In Eq.\ref{eq5}, $\sum_{v_j\in N(v_i)} p$ is the number of the
potentially influenced 1-hop neighbors, and $\sum_{v_j\in N(v_i)} outd(v_j)p^2-\sum_{ v_{u},v_{k}\in N(v_i),e_{uk}\in E} p^3$ is the number of the
potentially influenced 2-hop neighbors.

\textbf {Cost-effective diffusion value (CEDV) for a node}\cite{Han2014Balanced}. Eq.\ref{eq5} does not considering the cost of selecting a seed node. With considering the cost, the cost-effective diffusion value is computed by:
      \begin{align}
      \label{eq6}
       ce_2(v_i)=\sigma_{2}(v_i)/ c(v_i),
     \end{align}
      where $c(v_i)$ is the cost to activate $v_i$, $\sigma_2(v_i)$ is the 2-hop $EDV$ of $v_i$ and is computed by Eq.\ref{eq5}.

As introduced above, the CEDV computed by Eq.\ref{eq6} considered up to 2-hop neighbors and the cost for selecting a seed, which is also computation efficient. As shown in Eq.\ref{bim} for the BIM problem, the sum of the costs of the seed nodes should not larger than the budget $B$, so $ce_2(v_i)$ is chosen as the influence indicator in this paper.

\section{\textbf{Boost SA for the BIM}}
The effects of the topology of a network are analyzed and a strategy for candidate seed set selection is introduced in Subsection 4.1. Details of the proposed Boost SA are introduced in Subsection 4.2. Two tricks of the Boost SA are summarized in Subsection 4.3.

\begin{figure}
	\begin{minipage}{0.23\textwidth}  
		\centerline{\fbox{\includegraphics[width=1.5in]{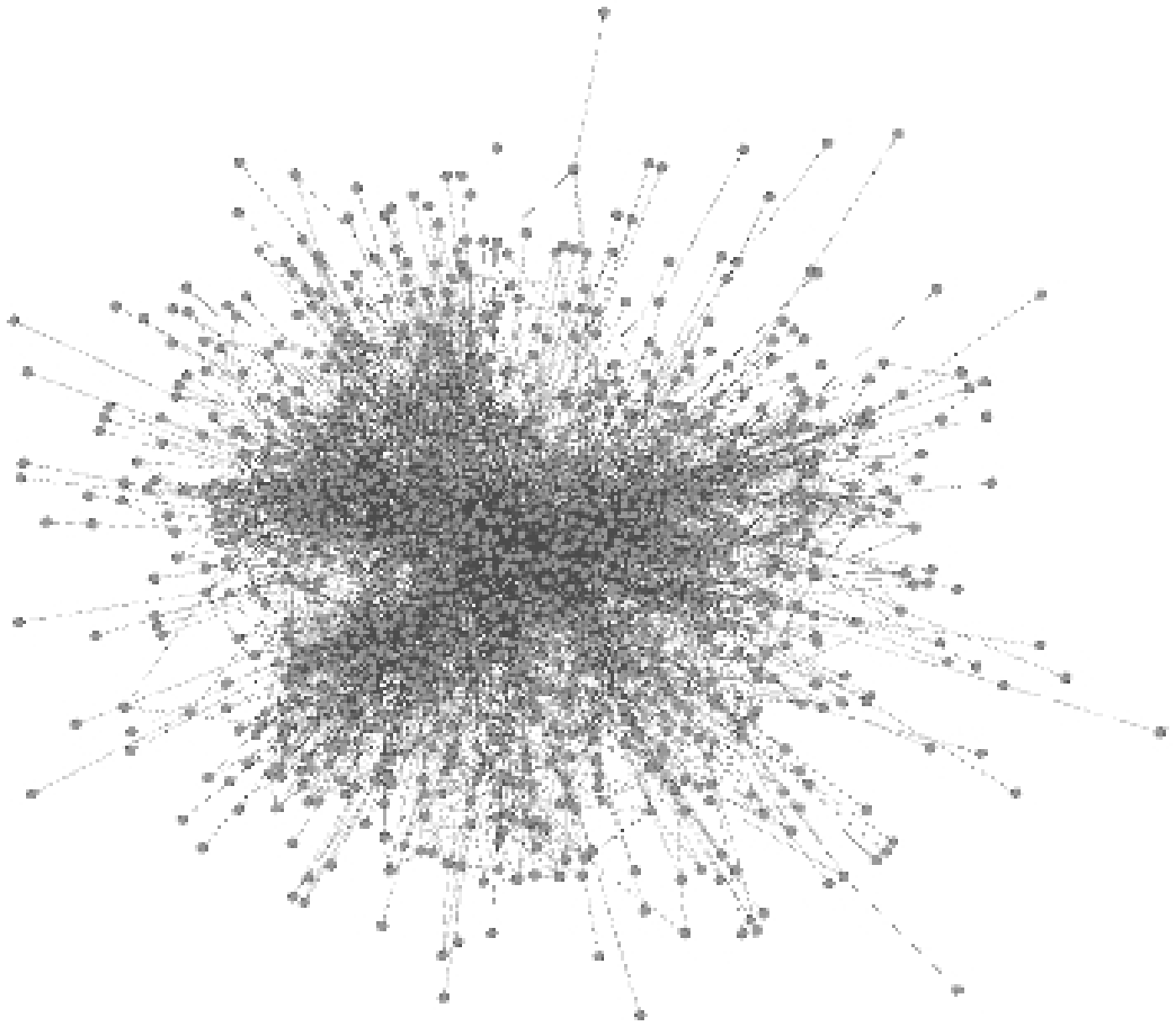}}}
		\caption{Structure of URV email network}
	\end{minipage}
	\hfill
	\begin{minipage}{0.23\textwidth}  
		\centerline{\fbox{\includegraphics[width=1.5in]{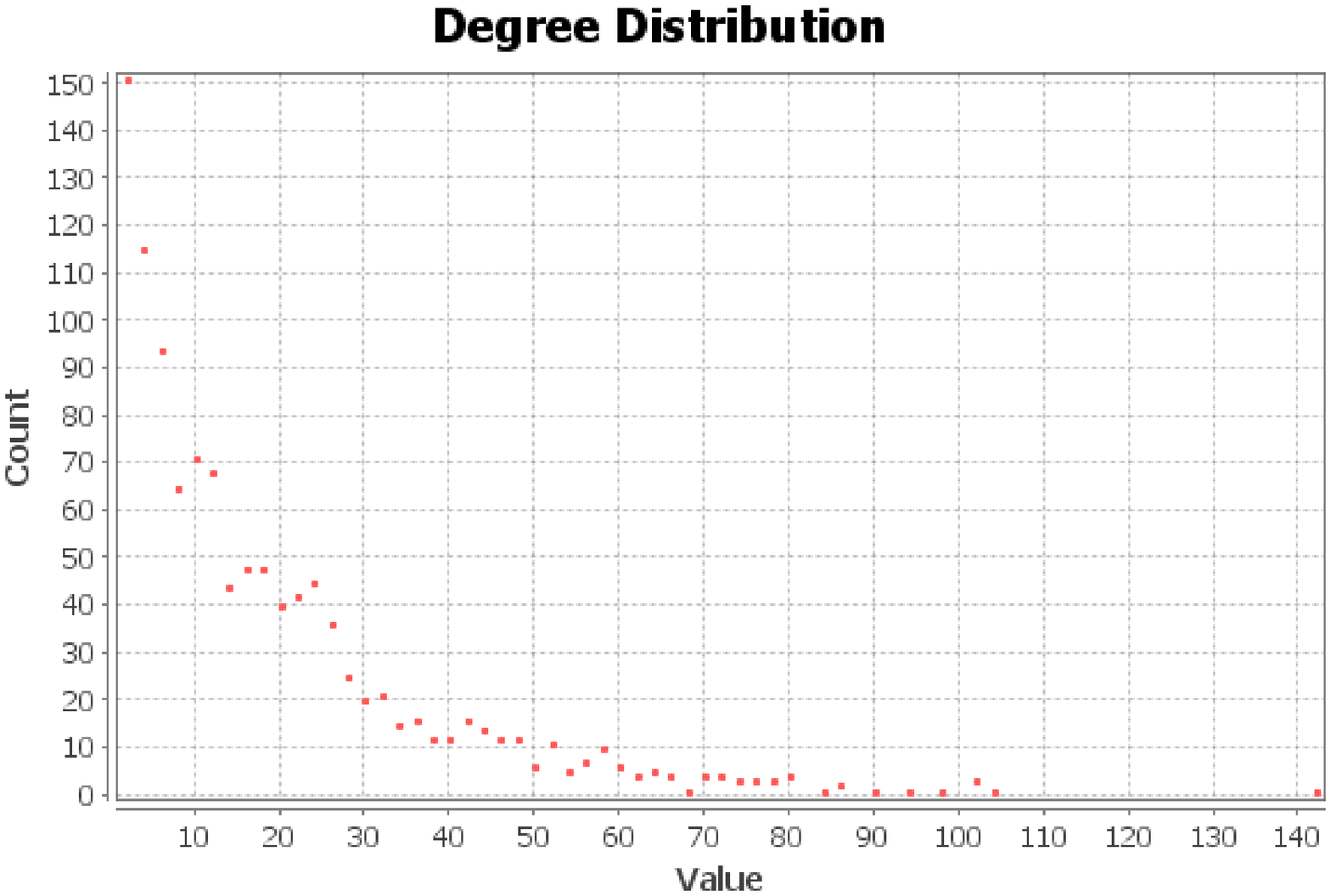}}}
		\caption{Degree Distribution of URV network}
		\label{fig3}
	\end{minipage}
\end{figure}

\begin{figure}
	\begin{minipage}{0.5\textwidth}
		\centerline{\includegraphics[width=2.2in]{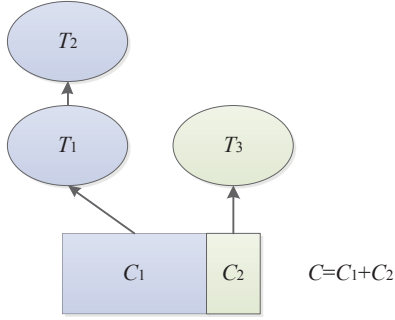}}
		\caption{The components of candidate set $C$}
		\label{fig4}
	\end{minipage}
\end{figure}

\subsection*{\textbf{4.1 Candidate seed set Selection}}
\indent As well known, different nodes have different influence usually. Based on statistical results of many real networks, the degree of nodes conforms to the power law distribution. Meanwhile, in many real scenarios, a little part of entities (nodes) usually take control of most of the resources. This interesting phenomenon is often summarized as ``2/8 rules'': about 20 percent of people (nodes) take control of 80 percent of treasure. As an example, Fig.2 is the structure of URV email network which contains 1133 nodes and 10903 edges. Corresponding to this network structure, Fig.3 is the degree distribution. It is obvious that the degree distribution follows a power law.

When considering the BIM problem by utilizing the 2/8 rules on the structure of a network, there is no doubt that the top 20\% nodes are of great value to achieve great influence spread. However, for the BIM, the node cost is about proportional to the number of out-degree, which means nodes with greater influence has more expensive cost. If we try to find out suitable seed set with cost-performance ratio, it is not a good choice to select nodes merely focus on its influence.

From Eq.\ref{eq6}, it is obvious that the indicator $ce_2(v_i)$ tends to find nodes with low degrees but have powerful neighbors. In other words, the indicator pays more attention to nodes whose neighbors contain more out-degrees rather than  the node itself has more direct neighbors. Since it's not economy to  select nodes with high degree centrality , as a concession, it is better to select their neighbors with low degrees so that the most influential nodes can be activated by the seeds.

For convenience, the set of top 20\% nodes with the maximum out-degree in the network is denoted as $T$, the set of bottom 80\% nodes with the minimum out-degree in the network is denoted as $H$.
In the Combination SA \cite{Han2014Balanced}, two candidate seed sets are proposed, which are called Billboard set ($S_b$) and Handbill set ($S_h$) respectively (see Section II.A for details).  $S_b$ is selected from $T$ and  $S_h$  is selected from $H$. Both $S_b$ and $S_h$ are selected with budget $B$. The final seed set $S$ is obtained by iteratively replacing each node in $S_b$ by a or several nodes in $S_h$ with less or equal cost. Although the main idea of this combination is to find cost-effective nodes by the replacement, but the separation of two candidate sets ignores the connections of nodes between the two sets, which will result in unnecessary budget waste.

 This paper proposes to use only one candidate set $C$ to reduce the solution space and utilize the connections between nodes, detailed flow chat is shown in Algorithm 1 and the components of $C$ are shown in Fig.\ref{fig4}.

\begin{algorithm}
	\caption{ Candidate seed set selection.}
	\begin{algorithmic}[1]
		\REQUIRE
		Graph $G (V, E)$, budget $B$, $\alpha, \beta$;\\
		\ENSURE
		Candidate seed set $C$;\\
		\FOR{each node $v_i \in V$}
		\STATE calculate  $outd(v_i)$;
		\ENDFOR
		\FOR{each node $v_i \in H$}
		\STATE calculate $ce_2(v_i)$ by $Eq.6$;
		\ENDFOR
		\STATE arrange the nodes in $H$ in descending order by $ce_2(v_i)$;
		\STATE $Total\_cost\leftarrow 0$,  $S\leftarrow \emptyset $, $C\leftarrow \emptyset $;
		\FOR{each node $v_i \in H$}
		\IF{$ce_2(v_i)+Total\_cost\leq \alpha B$}
		\IF{no node in $N_{out}(v_i)$ already exists in set $C$}
		\STATE $Total\_cost\leftarrow Total\_cost+ce_2(v_i)$;
		\STATE $C\leftarrow C\bigcup v_i $;
		\ENDIF
		\ENDIF
		\ENDFOR
		\FOR{each $v_t \in T$}
		\IF{$v_t$ has no in-degree neighbor in $C$}
		\STATE /*no direct parent node is selected*/
		\IF{$v_t$ has no in-degree neighbor in $T$ which has in-degree neighbor in $C$}
		\STATE /* no 2-hop parent node is selected*/
		\STATE find the in-degree neighbors $v_{in}(t)$ of node $v_t$
		\IF{$v_{in}(t)$ belongs to top $\beta$\% of set $H$}
		\STATE $C\leftarrow C\bigcup v_{in}(t)$;\\
		\ENDIF
		\ENDIF
		\ENDIF
		\ENDFOR
		\RETURN set $C$;
    \label{alg1}
	\end{algorithmic}
\end{algorithm}

Algorithm 1 calculates the cost-effective value $ce_2(v_i)$ of $v_i$  in $H$ by Eq.\ref{eq6} and arrange them in descending order (lines: 4-7). Select $v_i$ as candidate seed node if the total budget is less than $\alpha B$ and the out-direction neighbors of $v_i$ are not in candidate set $C$ (lines: 9-16)($C_1$ in Fig.\ref{fig4}). When the budget runs out,  check node $v_t \in T$, if $v_t$ has no  1-hop and 2-hop in-degree neighbor in  $C (C_1)$, check all its neighbors and find out the nodes belongs to top $\beta \%$ of $H$, put these nodes into set $C$ (lines: 17-28)($C_2$ in Fig.\ref{fig4}). Here, $\alpha$ and $\beta$ are two parameters to control the search space.

Instead of selecting influential nodes in $T$, the main idea of Algorithm 1 is selecting cost-effective nodes in $H$ to activate influential but costly nodes in $T$. As shown in  Fig.\ref{fig4}, candidate set $C$ composed of two parts $C_1$ and $C_2$.   When select nodes from $H$ for $C_1$, the total budget is $\alpha  B$ instead of $B$, where $1< \alpha < 2$, which controls the solution space. In the experiments of this paper, $\alpha  =1.5$. Too small value of $\alpha$, e.g., $\alpha=1$ (selecting candidate seed set with budget $B$),  will loss too much potential optimal nodes. 

When select nodes from $H$ for $C_2$, Algorithm 1 checks all nodes in $T$ and finds out nodes which can't be reached in 1-hop or 2-hop paths from a node in $C_1$. As shown in Fig.4, some nodes in set $T$ (denoted as $(T_1)$) can be directly reached by nodes in $C_1$. Some other nodes in set $T$ (denoted as $T_2$) can be reached by nodes in $T_1$, thus can be  reached by nodes in  $C_1$ within 2 hops. Apart from $T_1$ and $T_2$, there are still many nodes in $ T$ (denoted as $T_3$), which are influential and costly. Experiments on three real world networks  are shown in Tables \ref{table2}-\ref{table4}, which indicate that with the increase of budget, more and more nodes in $T$ can be reached, but there are still some nodes can't be reached within 2-hops.  Find a neighbor for each node of $T_3$  in $H$ which have promising value of $ce_2$  and add it into $C$ (line 20-26), denoted as $C_2$ in Fig.\ref{fig4}. In order to activate the nodes in $T_3$ as much as possible and select cost-effective nodes as candidate seeds, we set $\beta\% = 60\%$.



\renewcommand{\algorithmicrequire}{ \textbf{Input:}} 
\renewcommand{\algorithmicensure}{ \textbf{Output:}} 

\textbf{Time Complexity of Algorithm 1.} To calculate out-degree of each node takes $O(n)$ time (line 1-3), where $n$ represents the number of network nodes. The cost-effective value calculation of nodes needs $O(n^{2})$ time (line 4-6); the arrangement of nodes in $H$ spends $O(n)$ time (line 7); nodes selection needs $O(kn)$ time (line 9-16);  check of $T$ and construct $C_2$ needs $O(n^{2})$ time (line 17-25). Therefore, the time complexity of Algorithm 1 is $O(n^{2})$.

When selection of the candidate seed set $C$ is finished, the next step of the proposed Boost SA is selecting nodes from $C$ for initial seed set $S$ (initialization).
\begin{table}[!htbp]
\centering
\caption{$T_1, T_2$, and $T_3$ in the URV email network}
\label{table2}
\begin{tabular}{cccc}
  \hline
  Budget & \tabincell{c}{$C$} & \tabincell{c}{$T_1$ \& $T_2$} & \tabincell{c}{$T_3$}\\
  \hline
  50 & 62 & 187 & 39 \\
  100 & 75 & 217 & 9 \\
  150 & 114 & 224 & 2 \\
  200 & 156 & 224 & 2 \\
  250 & 207 & 225 & 1 \\
  300 & 262 & 225 & 1 \\
  \hline
\end{tabular}
\end{table}
\begin{table}[!htbp]
\centering
\caption{$T_1, T_2$, and $T_3$ in the Wiki-Vote network}
\label{table3}
\begin{tabular}{cccc}
  \hline
  Budget & \tabincell{c}{$C$} & \tabincell{c}{$T_1$ \& $T_2$} & \tabincell{c}{$T_3$}\\
  \hline
  1000 & 901 & 450 & 1209 \\
  1500 & 1237 & 541 & 1118 \\
  2000 & 1658 & 564 & 1095 \\
  2500 & 2179 & 602 & 1057 \\
  3000 & 2787 & 653 & 1006 \\
  3500 & 3426 & 682 & 977 \\
  4000 & 4118 & 687 & 972 \\
  \hline
\end{tabular}
\end{table}

\begin{table}[!htbp]
\centering
\caption{$T_1, T_2$, and $T_3$ in the NetHEPT network}
\label{table4}
\begin{tabular}{cccc}
  \hline
  Budget & \tabincell{c}{$C$} & \tabincell{c}{$T_1$ \& $T_2$} & \tabincell{c}{$T_3$}\\
  \hline
  1000 & 941 & 1295 & 1106 \\
  1500 & 1354 & 1337 & 1064 \\
  2000 & 1840 & 1350 & 1051 \\
  2500 & 2393 & 1359 & 1042 \\
  3000 & 2995 & 1359 & 1042 \\
  3500 & 3625 & 1359 & 1042 \\
  4000 & 4173 & 1359 & 1042 \\
  \hline
\end{tabular}
\end{table}

\subsection*{\textbf{4.2 The proposed Boost SA}}

The proposed Boost SA is based on the idea of simulated annealing.  The random initialization strategy is used to produce the initial seed set. The pseudocode of initialization is shown in \textbf{Algorithm \ref{alg:initial}}. At first, $k$ initial seed sets $S_i$, $i=1,..., k$, are randomly selected from $C$, each with budget $B$ (line 2). For each $S_i$, a node in $S_i$ is randomly replaced to produce its neighbor set (line 7), and the optimal solution is selected based on the idea of simulated annealing algorithm. That is to choose the optimal solution as much as possible, and select the sub-optimal solution with a certain probability to jump out of the local optimal solution (line 9-16). At the same time, the number of times each node is selected is recorded (line 18). Finally, the initial seed set $S$ is obtained according to the number of times each node in $C$ is selected (line 20-21).

The pseudocode of Boost SA is shown in \textbf{Algorithm \ref{alg:BSA}}. As shown in Algorithm \ref{alg:BSA}, from the candidate seed set $C$, Boost SA has two stages of iteration. The outer iteration is controlled by the initial temperature $t_0$ and the threshold temperature $t_f$. In outer iteration, we create $GP$ groups of $S$ and each group will go through the inner iteration.

The inner iteration is controlled by the number of total groups $GP$ and the number of loops $q$. In inner iteration,  randomly select a node in set $S$ and replace it with the equal (or less) costs of nodes in $C$. If the influence difference is bigger than zero, accept it; otherwise, create a random value in $(0,1)$ and compare it with $exp(-\Delta E / T_t) $ and decide whether to accept or reject the replacement. It is worth pointing out that this comparison with random value allows to accept a worse solution, which can help of jumping out of local optimal and find a better solution. The replacement above will be performed for $q$ time. Calculate the number of times that each nodes are selected in all $GP$ groups, which is a nodes list, denoted as $L(C)$. In $L(C)$, the nodes are sorted in descending order for selecting nodes for $S$ (see Algorithm \ref{alg:vote}).

 We record the influence value of set $S$ per iteration in $max\_infulence$. If the value of  $max\_infulence$ has repeated for $num$ times, the iteration is stopped. Then  break the iteration and return the final set $S$ (line 22-24 in Algorithm \ref{alg:BSA}).


\begin{algorithm}[!htbp]
	\caption{Initialization.}
	 \label{alg:initial}
	\begin{algorithmic}[1]
		\REQUIRE
		Graph $G (V, E)$, budget $B$, candidate seed set $C$,  initial temperature $t_0$, the number of loop $q$, the number of $k$.\\
		\ENSURE
		The initial seed set $S$;\\
		\STATE $temp\_q\leftarrow q $;
		\STATE Randomly generate $k$ initial seed sets  $S_i$ from $C$, $i=1,\cdots,k$ , $S_i$ has budget $B$ ;
		\FOR{$i=1,\cdots,k$ of $S_i$}
		\STATE $q\leftarrow temp\_q$, $flag\leftarrow false $;
		\WHILE{$q>0$}
		\STATE $q\leftarrow q-1$;
		\STATE Randomly replacing a node in $S_i$ by nodes in $C$ with less than or equal cost, formulate a neighbor set $S_i^{'}$ ;
		\STATE Compute the influence difference $\Delta E\leftarrow \sigma(S_i^{'})-\sigma(S_i)$;
		\IF {$\Delta E>0$}
		\STATE $S_i\leftarrow S_i^{'}$, $flag\leftarrow true$;
		\ELSE
		\STATE create a random number $\varepsilon \in U(0,1)$;
		\IF {$exp(\Delta E /t_0)>\varepsilon$}
		\STATE $S_i\leftarrow S_i^{'}$;
		\ENDIF
		\ENDIF
		\ENDWHILE
		\STATE Record the number of times that each node in set $C$ is selected, denoted as $L(C)$;
		\ENDFOR
		\STATE Let nodes in set $C$ are sorted in descending order of the selected number;
		\STATE Select nodes in turn from set $C$  into set $S$ with budget $B$;
		\RETURN initial seed set $S$;
	\end{algorithmic}
\end{algorithm}


\begin{algorithm}[!htbp]
	\caption{ Seed set selection based on Simulated Annealing.}
	\label{alg:BSA}
	\begin{algorithmic}[1]
		\REQUIRE
		Graph $G (V, E)$, budget $B$, candidate seed set $C$, initial temperature $t_0$, temperature drop $\Delta T$, Threshold temperature $t_f$, Objective function $\sigma(C)$, the number of loop $q$, the number of group $GP$ and set number $num$, initial seed set $S$.\\
		\ENSURE
		The final seed set $S$;\\
		\STATE $ T_t\leftarrow t_0 $, $temp\_q\leftarrow q $;
		\STATE $S\leftarrow $ initial seed set $S $;
		\WHILE{$T_t>t_f$}
		\FOR{$i=1,\cdots,GP$}
		\STATE $q\leftarrow temp\_q$, $flag\leftarrow false $;
		\WHILE{$q>0$}
		\STATE $q\leftarrow q-1$;
		\STATE Randomly replacing a node in $S$ by nodes in $C$ with less than or equal cost, formulate a neighbor set $S^{'}$ ;
		\STATE Compute the influence difference $\Delta E\leftarrow \sigma(S^{'})-\sigma(S)$;
		\IF {$\Delta E>0$}
		\STATE $S\leftarrow S^{'}$, $flag\leftarrow true$;
		\ELSE
		\STATE create a random number $\varepsilon \in U(0,1)$;
		\IF {$exp(\Delta E /T_t)>\varepsilon$}
		\STATE $S_i\leftarrow S^{'}$;
		\ENDIF
		\ENDIF
		\ENDWHILE
		\STATE Record the number of times that each node in set $C$ is selected, denoted as $L(C)$;
		\ENDFOR
		\STATE Update $S$ and $max\_influence$ by Algorithm \ref{alg:vote};
		\IF{the value of $max\_influence$ is repeated for $num$ times}
		\STATE Break;
		\ENDIF
		\STATE $T_t \leftarrow T_t-\Delta T$;
		\ENDWHILE
		\RETURN set $S$;
	\end{algorithmic}
\end{algorithm}

\begin{algorithm}[!htbp]
  \caption{ Voting for new set S.}
  \label{alg:vote}
 
  \begin{algorithmic}[1]
    \REQUIRE
        $L(C)$.\\
    \ENSURE
       $S$ and $max\_influence$.\\
    \STATE Let nodes in set $C$ are sorted in descending order of the selected number;
    \STATE $temp\_S \leftarrow S$, $S \leftarrow \emptyset$;
    \STATE Select nodes in turn from set $C$ to set $S$ with the budget $B$;
    \IF{$\sigma(S)-\sigma(temp\_S) > 0$}
    \STATE  $max\_influence \leftarrow \sigma(S)$ ;
    \ELSE
    \STATE $S\leftarrow temp\_S$;
    \STATE $max\_influence \leftarrow \sigma(temp\_S)$ ;
    \ENDIF
    \RETURN $S$ and $max\_influence$;
  \end{algorithmic}
\end{algorithm}

\subsection*{\textbf{4.3 Two Tricks for Optimization}}
\textbf{Trick 1. Voting for new seed set S.} The computation of traditional Greedy Algorithm is expensive. In order to cut down the running time, SA Algorithm is utilized to replace it. As described above, the main structure of Combination SA Algorithm \cite{Han2014Balanced} is two stages of iteration. The outer iteration is controlled by the number of nodes in Billboard set and the inner iteration is controlled by constant value $q$. In essence, Combination SA performs millions of iterations just in order to fully utilize the replacement operation and obtain reliable results. Random replacement is introduced to strengthen the ability to jump out of local optimum. However, this trick will also bring in uncertainty which has a negative influence on final solution due to the fact that set $S$ is a whole group.

As a popular theory in machine learning, Ensemble Learning has been widely used in many famous algorithms (e.g. KNN \cite{Hart2003The} and Random Forests \cite{Ho1995Random}). The main idea of this theory is quite simple: different methodes has different performance for the same problem, may be none of them can get the best result. However, if we collect those strategies and make the final decision, we can achieve a better result than just adopt a single strategy. This phenomenon is often called ``the wisdom of crowds''.

For SA method in BIM problem, we utilize the idea of Ensemble Learning and generate $GP$ sets $S$ for inner iteration. In Algorithm 4, nodes in $C$ are sorted in descending order of the recorded selected number (line 1). It makes sense to choose nodes that have been selected in many groups. The record and reselect operation is just like voting for the promising nodes. Then select nodes into set $S$ with budget $B$. After that,  compare new influence $\sigma(S)$ with $\sigma(temp\_S)$ in last iteration (line 4 in Algorithm 4). The comparison reflects the belief that after $GP$ groups' independent iterations and voting process, the result is reliable. If new set $S$ can't provide better solution, we give up the result of this outer iteration in order to keep the temp optimal result (line 7-8 in Algorithm 4).


\textbf{Trick 2. Adaptive interrupt mechanism.} As analysis in motivation 1, the Combination SA contains two stages of iteration. The outer iteration is controlled by the number of nodes $K$ in Billboard set, and the inner iteration is controlled by constant value $q$, which means the total number of iterations is $K*q$. As the budget $B$ grows, the number of outer iteration keeps on growing and the runtime is proportional to the budget increase. This property has a negative effect on practical application.

Experiments are done to show the influence spread computed by Eq.5 with different budgets, the results on Wiki-Vote network are shown in Fig.5, which shows that the influence spread  remains unchanged after several iterations, more number of outer iteration is only a waste of running time. Similar results are found in the URV email and NetHEPT datasets, which are not shown here.  Inspired by this observation, we record influential value $max\_influence$ of set $S$ in each iteration and compare it with previous $num$ influential values $max\_influence$. If the past $num$ $max\_influence$ equal to the latest  $max\_influence$, the algorithm believes further iteration has out of action and break the iteration. Runtime comparison in next section will verify the advantage of this adaptive interrupt strategy.

\textbf{Time Complexity.}  The initialization by Algorithm \ref{alg:initial} takes $O(k*q*n)$ time. In Algorithm \ref{alg:BSA}, the outer iteration is controlled by the value of $GP$ and $(t_0-t_f)/\Delta T$. The inner iteration is controlled by value $q$. In inner iteration, the calculation of influence indicator in line 9 takes $O(n)$ time. Voting for new set $S$ by Algorithm 4 (line 21 in Algorithm \ref{alg:BSA}) takes $O(n)$ time. Therefore, the running time of Algorithm \ref{alg:BSA} is $O((t_0-t_f)*GP*q*n/\Delta T) \approx O(n^2)$. So the time complexity of the Boost SA is $O(n^2)$.

\section{\textbf{Experiments}}

\subsection{\textbf{Network Datasets}}
Three real world networks and four synthetic networks are used in experiments. The parameters of the three real world datasets are summarized in Table \ref{table5}:

URV email network \cite{Guimer2003Self}. The first dataset is a small size network, which is generated using email interchanges between members of the University Rovira i Virgili with totally 1133 nodes and 10903 edges. There is an edge $e_{ij}$ in the network if person $v_{i}$ send or receive from person $v_{j}$ at least one email.

Wiki-Vote network \cite{Leskovec2010Signed}. The second dataset is a middle size network, which is generated from Wikipedia community with totally 7115 nodes and 103689 edges. In this network, each node represents a user and a directed edge from node $v_i$ to $v_j$ represents that user $v_i$ voted for user $v_j$.

NetHEPT network \cite{Chen2009Efficient}. The third dataset is a large size network, which is a collaboration network between authors in high energy field and has been widely used as a baseline network. It has 12008 nodes and 118521 edges. In this network, a node represents an author and an edge between two nodes represents the cooperation of a paper between the two authors. The data in network covers papers in the period from January 1993 to April 2003 (124 months).

\begin{table}[!htbp]
\centering
\caption{Statistics of three real networks}
\label{table5}
\begin{tabular}{cccc}
  \hline
  Network & Nodes & Edges & Average Degree\\
  \hline
  URV email & 1133 & 10903 & 9.62 \\
  Wiki-Vote & 7115 & 103689 & 14.57 \\
  NetHEPT & 12008 & 237010 & 19.74 \\
  \hline
\end{tabular}
\end{table}

The synthetic networks are generated by benchmark proposed by \cite{Andrea2009Benchmarks}. It can generate weighted community networks with power-law distribution of node degree and many other topological structure parameters. The four networks share several parameters listed in Table \ref{table6},  the number of nodes are 2000, 5000, 10000 and 30000 respectively.

\begin{table}[!htbp]
\centering
\caption{Detailed parameters for synthetic networks}
\label{table6}
\begin{tabular}{cccc}
  \hline
  Definition & Symbol & Value\\
  \hline
  Number of vertices& N & 2000/5000/10000/30000\\
  Average Degree& avg-D & 5/10/10/10\\
  Maximum Degree& max-D & 50\\
  Mixing parameter& $\mu$ & 0.15 \\
  \tabincell{c}{Minimum for the\\ community sizes}& min-c & 20 \\
  \tabincell{c}{Maximum for the\\ community sizes}& max-c & 50 \\
  Exponent for the \\degree distribution & Exp-D & 2 \\
  Exponent for the \\community size distribution & Exp-C & 1 \\
  \hline
\end{tabular}
\end{table}

\subsection{\textbf{Comparing Algorithms}}

\textbf{MODPSO.} The MODPSO algorithm \cite{Yang2018Influence} is a modified particle swarm optimization method. It considers the BIM problem as a multi-objective optimization problem and try to maximize the number of activated nodes with least budget. As a novel method, it's persuasive to make a comparison with it.

\textbf{CELF.} The CELF algorithm \cite{Leskovec2007Cost} is an improvement on the greedy algorithm. It can speed up the latter for 700 times. CELF has been the baseline method for IM problem. Since the algorithm is designed for IM problem, we make some changes so that it can satisfy the BIM problem.

\textbf{MaxDegree.} The MaxDegree algorithm \cite{Tardos2003Maximizing} is a heuristic method that can be easy to come up with. In this algorithm, degree is the only indicator to calculate the ability of influence spread. It sorts nodes in descending order and selects nodes with budget $B$.

\textbf{Combination SA.} As introduced in Section II, the Combination SA  \cite{Han2014Balanced} is the first algorithm that utilizes simulated annealing method to deal with the BIM problem. 

\textbf{Boost SA.} The proposed algorithm as described in Algorithm 3,  which is an improvement over the Combination SA algorithm.

The IC model is adopted for all experiments. In order to keep the accuracy of different algorithms, we adopt simulated method \cite{Tardos2003Maximizing} to compute the final influence spread of selected seed set of each algorithm. 10000 times of Monte-Carlo simulation are performed and we take the average influence spread as the expected result. In order to keep the reliability, each algorithm is independently run 30 times on each dataset. All experiments are performed on a computer with 2.10GHz AMD Ryzen5 and 16G memory. All algorithms are written in MATLAB R2017a.

\begin{figure}[!htbp]
	
	\begin{minipage}{0.24\textwidth}  
		\centerline{\includegraphics[width=1.7in]{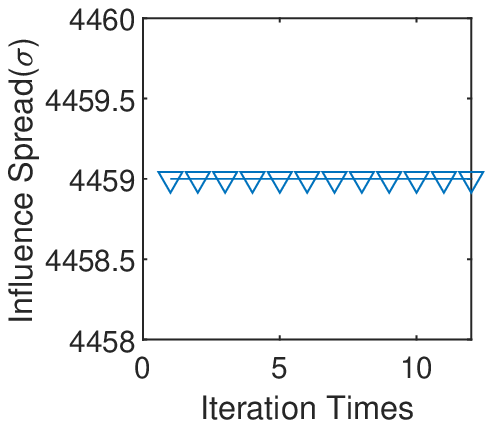}}
		\centerline{(a).B = 1000}
	\end{minipage}
	\hfill
	\begin{minipage}{0.24\textwidth}  
		\centerline{\includegraphics[width=1.7in]{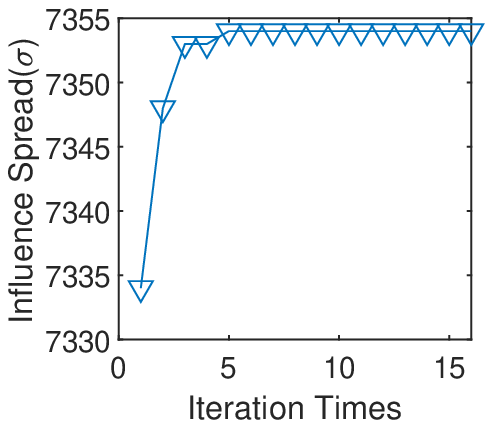}}
		\centerline{(b).B = 2000}
	\end{minipage}
	\vfill
	\begin{minipage}{0.24\textwidth}  
		\centerline{\includegraphics[width=1.7in]{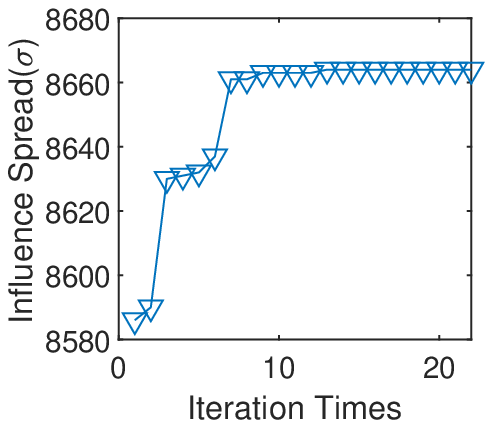}}
		\centerline{(c).B = 3000}
	\end{minipage}
	\hfill
	\begin{minipage}{0.24\textwidth}
		\centerline{\includegraphics[width=1.7in]{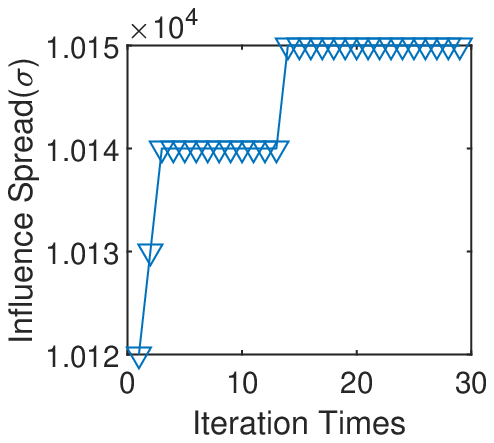}}
		\centerline{(d). B = 4000}
	\end{minipage}
	\caption{Influence spread on Wiki-Vote with different budgets}
\end{figure}

\subsection{\textbf{Parameter Setting}}
For Combination SA algorithm, we directly utilize the settings of parameters in Ref \cite{Han2014Balanced}: $t_0=1000000$, $q=1000$, $\Delta T=1000$ and $t_f=100000$.

For convenience in comparison between Combination SA and the proposed Boost SA,  the same parameter settings are used in the Boost SA. Besides, we set $k=10$, $GP=3$, and $num=10$. The cost of a node $v_i$ is defined as
     \begin{equation}
      \label{eq7}
      c(v_i)=outD(v_i)*p+1,      
      \end{equation} 
where $outD(v_i)$ is the out-degree of $v_i$ and $p=0.1$ is the propagation probability.

\subsection{\textbf{Experiments on real networks}}

\begin{figure}
	
	\begin{minipage}{0.24\textwidth}  
		\centerline{\includegraphics[width=1.7in]{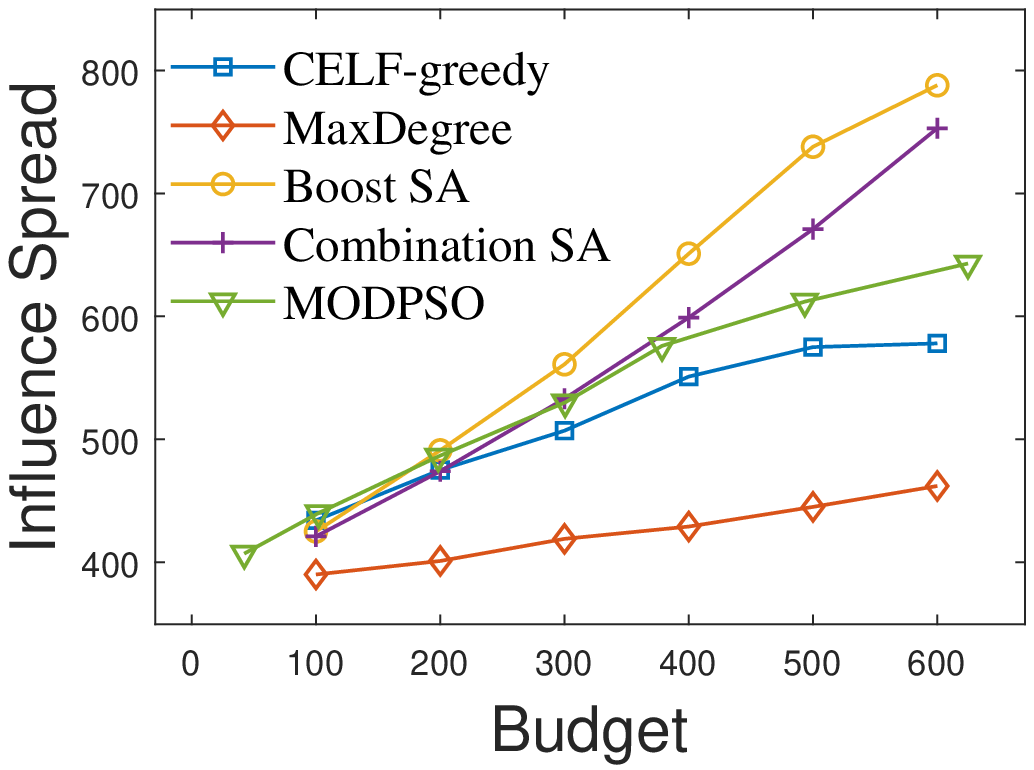}}
		\centerline{(a).URV email network}
	\end{minipage}
	\hfill
	\begin{minipage}{0.22\textwidth}  
		\centerline{\includegraphics[width=1.7in]{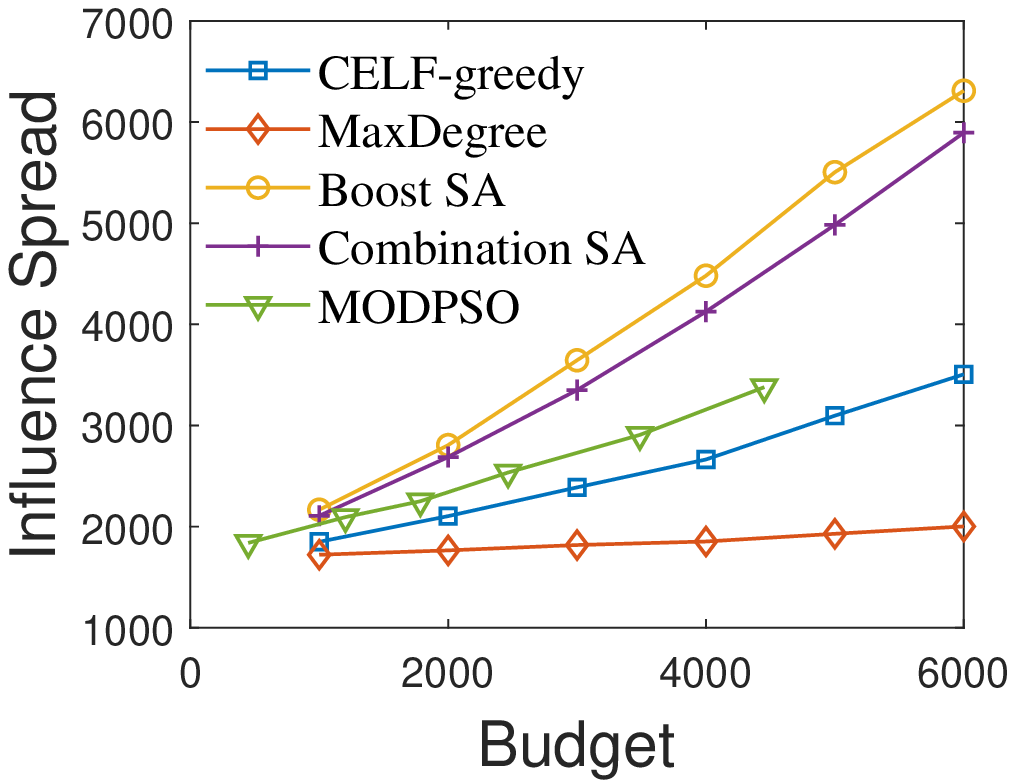}}
		\centerline{(b).Wiki-Vote network}
	\end{minipage}
	\vfill
	\begin{minipage}{0.24\textwidth}  
		\centerline{\includegraphics[width=1.70in]{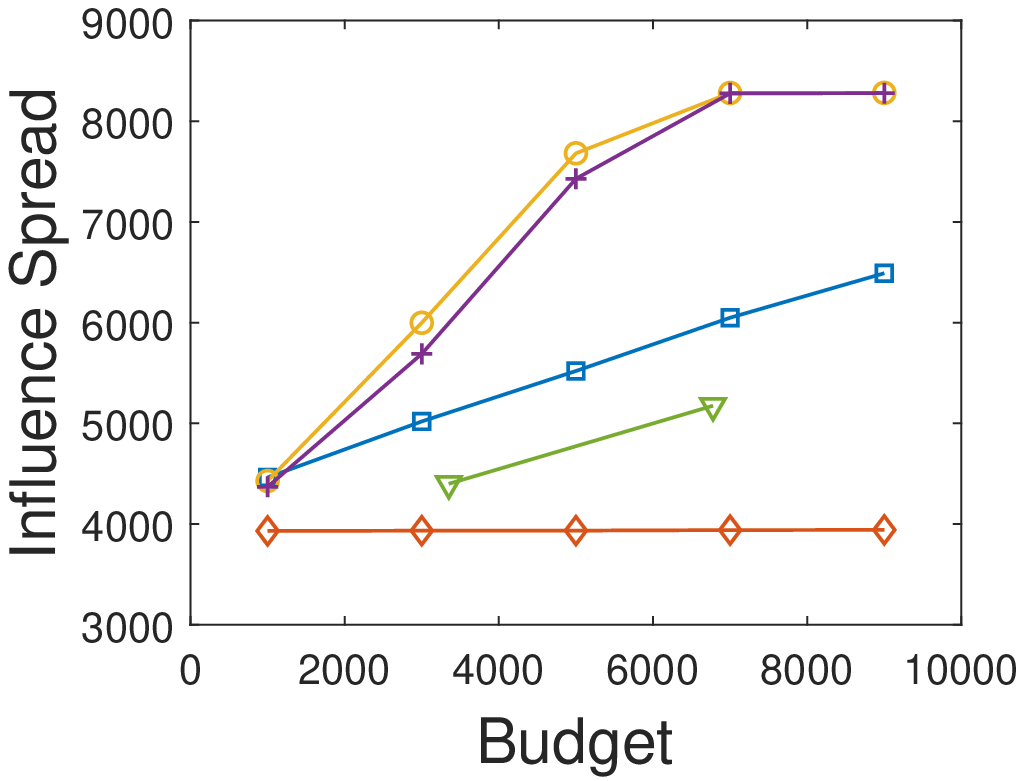}}
		\centerline{(c).NetHEPT network}
	\end{minipage}
	\hfill
	\begin{minipage}{0.22\textwidth}
		\centerline{\includegraphics[width=1.7in]{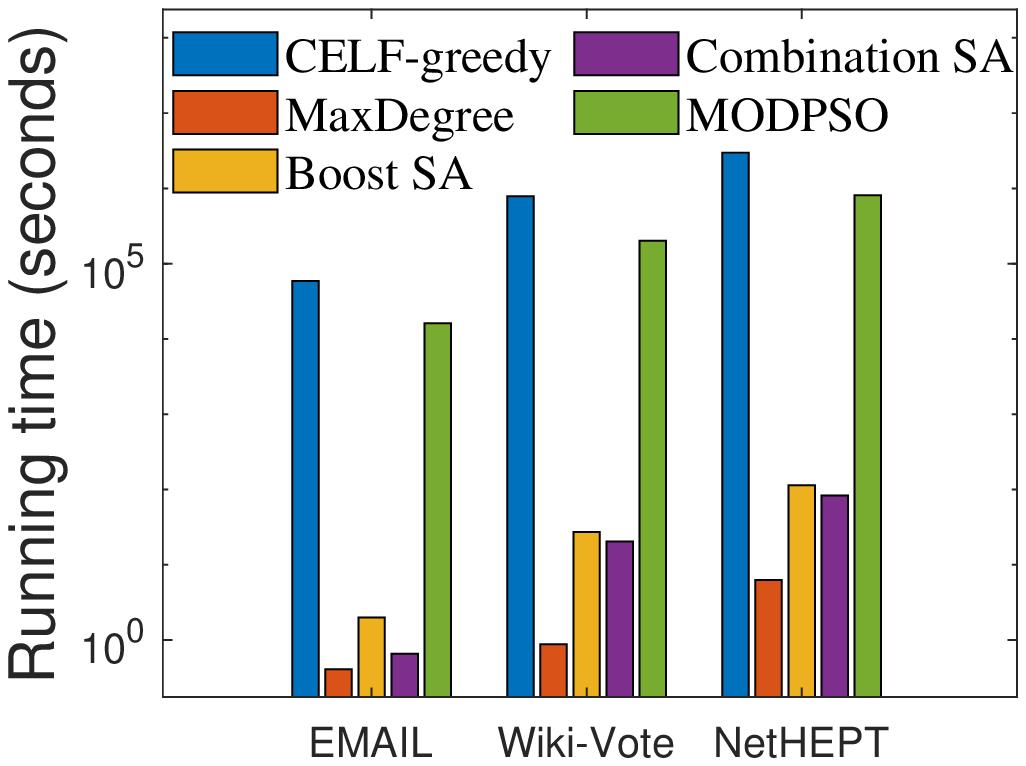}}
		\centerline{(d).Running time}
	\end{minipage}
	\caption{Influence spread on real world networks}
	\label{fig6}
\end{figure}

The experimental results are shown in Fig.\ref{fig6} with $q=1000$ for both Boost SA and Combination SA. It can be seen that in most cases the Boost SA obtains the best seed set $S$ with different budgets, which has the greatest influence.

More specifically, for the URV email network, when $150<B<300$, Boost SA is better than both Combination SA and MODPSO. When $B>300$, the advantage of Boost SA over other algorithms becomes more and more obvious.  When $B = 500$, Boost SA is 9.90\% better than Combination SA. In overall, Boost SA is 4\% better than Combination SA algorithm from $B$ = 100 to 600 in average.

For Wiki-Vote network, the Boost SA always obtains the best seed set $S$. As the increasing of Budget from 1000 to 6000, the advantage of the Boost SA is more obvious than the Combination SA and other algorithms. Boost SA outperforms Combination SA about 7.5\% in average.

For NetHEPT network, Boost SA obviously performs better than the Combination SA and other algorithms when $B<7000$. Boost SA is 3.41\% better than Combination SA in average. When $B>7000$, Boost SA almost equivalent to Combination SA. Due to expensive time consuming, we only run MODPSO at two budget points. The results reflect that Boost SA performs better than MODPSO with very little running time.

Fig.\ref{fig6}(d) shows the running time of the algorithms on three real networks when $B=100$. As we know, the greedy algorithm CELF consumes the most time. The running time of Boost SA and Combination SA are close together.

For a comprehensive comparison between the Boost SA and Combination SA for different values of $B$ and $q$, we set parameter $q$ as 1000, 500, 100 in Boost SA and Combination SA for different budgets $B$ on different datasets. The experimental results are shown in Tables \ref{table7},\ref{table8} and \ref{table9}, which indicate the influence value increases gradually with the increase of $q$. 

With equal number of $q$, the influence spread of the Boost SA outperforms Combination SA obviously, the running time of Boost SA is approximate to that of Combination SA. Comparing $q=100$ of Boost SA with $q=1000$ of Combination SA, the inflence  spread of the Boost SA outperforms Combination SA obviously also, the corresponding running time of Boost SA are all less than the Combination SA. In summary, the Boost SA can outperform Combination SA in influence spread with less running time.

\begin{table}[H]
	
	\centering
	\caption{Influence/running time on URV email network}
	\label{table7}
	\begin{tabular}{|c|c|c|c|}
		\hline
		Budget&\multicolumn{3}{c|}{\tabincell{c}{Boost SA\\(influence/running time)}}\\
		\hline
		&q=1000&q=500&q=100\\
		\hline
		100 &425/1.729 &425/1.237 &424/0.955\\
		\hline
		200 &490/2.908 &488/1.574 &487/1.107\\
		\hline
		300 &563/2.471 &560/1.824 &556/1.280\\
		\hline
		400 &656/3.121 &656/2.229 &654/1.554\\
		\hline
		500 &740/3.399 &737/2.230 &736/1.572\\
		\hline
		600 &782/3.443 &782/2.304 &781/1.688\\
		\hline
		Budget&\multicolumn{3}{c|}{\tabincell{c}{Combination SA\\(influence/running time)}}\\
		\hline
		&q=1000&q=500&q=100\\
		\hline
		100 &421/1.036 &421/0.948 &417/0.854\\
		\hline
		200 &475/1.345 &473/1.168 &471/0.925\\
		\hline
		300 &530/1.515 &529/1.252 &529/1.039\\
		\hline
		400 &600/1.797 &597/1.412 &595/1.101\\
		\hline
		500 &674/1.970 &671/1.514 &670/1.167\\
		\hline
		600 &753/2.272 &753/1.653 &752/1.167\\
		\hline
	\end{tabular}
\end{table}



\begin{table}[H]

	\centering
	\caption{Influence/running time on Wiki-Vote network}
	\label{table8}
	\begin{tabular}{|c|c|c|c|}
		\hline
		Budget&\multicolumn{3}{c|}{\tabincell{c}{Boost SA\\(influence/running time)}}\\
		\hline
		&q=1000&q=500&q=100\\
		\hline
		500  &1898/19.402 &1897/19.312 &1891/18.911\\
		\hline
		1000 &2169/21.668 &2164/20.591 &2163/19.571\\
		\hline
		1500 &2474/26.445 &2473/23.147 &2473/20.643\\
		\hline
		2000 &2817/33.517 &2817/26.860 &2816/21.558\\
		\hline
		2500 &3210/45.485 &3210/32.940 &3202/23.181\\
		\hline
		3000 &3651/53.909 &3650/42.984 &3646/25.892\\
		\hline
		Budget&\multicolumn{3}{c|}{\tabincell{c}{Combination SA\\(influence/running time)}}\\
		\hline
		&q=1000&q=500&q=100\\
		\hline
		500  &1866/25.176 &1866/22.687 &1866/20.966\\
		\hline
		1000 &2108/26.536 &2108/23.903 &2108/21.737\\
		\hline
		1500 &2389/28.230 &2389/24.581 &2388/23.502\\
		\hline
		2000 &2689/30.909 &2688/26.169 &2688/24.985\\
		\hline
		2500 &3004/35.416 &3004/30.684 &3001/27.468\\
		\hline
		3000 &3351/40.297 &3351/31.694 &3350/30.329\\
		\hline
	\end{tabular}
\end{table}

\begin{table}[H]
	\centering
	\caption{Influence/runtime on NetHEPT network}
	\label{table9}
	\begin{tabular}{|c|c|c|c|}
		\hline
		Budget&\multicolumn{3}{c|}{\tabincell{c}{Boost SA\\(influence/runtime)}}\\
		\hline
		&q=1000&q=500&q=100\\
		\hline
		1000 &4435/201.030 &4434/197.661 &4431/188.314\\
		\hline
		2000 &5121/215.250 &5119/208.960 &5117/206.970\\
		\hline
		3000 &6005/232.090 &6003/219.880 &6002/215.020\\
		\hline
		4000 &6933/240.230 &6929/231.190 &6925/227.980\\
		\hline
		5000 &7636/251.730 &7632/241.260 &7620/243.170\\
		\hline
		Budget&\multicolumn{3}{c|}{\tabincell{c}{Combination SA\\(influence/runtime)}}\\
		\hline
		&q=1000&q=500&q=100\\
		\hline
		1000 &4368/197.806 &4368/192.814 &4367/189.033\\
		\hline
		2000 &4952/217.020 &4951/209.560 &4949/197.590\\
		\hline
		3000 &5692/252.090 &5690/228.030 &5689/207.850\\
		\hline
		4000 &6519/274.770 &6518/252.120 &6518/221.080\\
		\hline
		5000 &7432/285.400 &7429/272.490 &7429/233.570\\
		\hline
	\end{tabular}
\end{table}

\subsection{\textbf{Experiments on synthetic networks}}

The experiment results of synthetic networks are shown in Fig.7. In this experiment,  $q=1000$ for both Boost SA and Combination SA. From to results of real networks and expensive runtime consuming, it is unnecessary to run MODPSO algorithm on synthetic networks. We just compare the rest four algorithms on synthetic networks.

For network-2000, Boost SA always performs the best for different budgets. Boost SA performs 5.68\% better than Combination SA in average. In addition, Boost SA and  Combination SA are always better than CELF and MaxDegree. When $B<300$, MaxDegree is slightly better than CELF.  When $B>300$, CELF is always better than MaxDegree.

For network-5000, when $B<1100$, CELF performs slightly better than Boost SA and Combination SA. When $1100<B<1500$, Boost SA is 5.31\% better than CELF and CELF is slightly better than Combination SA. Since $B>1500$, Boost SA is 0.65\% better than Combination SA and Combination SA is 34.30\% better than CELF. MaxDegree performs worse than other three algorithms with different budgets.

For network-10000, Boost SA always performs the best for different budgets. When $B<1500$, CELF performs slightly better than Combination SA, but when $B>1500$, Combination SA is better than CELF and the advantage gradually expands to about 10.10\% when $B=4000$. Boost SA performs about 3.84\% better than Combination SA in average with different budgets.

For network-30000, Boost SA always performs the best for different budgets. In particular, Boost SA performs 5.77\% better than Combination SA. Combination SA performs almost as well as CELF and Combination SA.

The running time on the four synthetic networks with $B=1000$ is shown in Fig.\ref{fig8}. With the increase of network size, the running time increases gradually. Specifically, CELF consumes the most the running time. Boost SA and Combination SA consume roughly the same running time. It shows that the proposed Boost SA can achieve better result with approximate running time.

\begin{figure}
\begin{minipage}{0.24\textwidth}
  \centerline{\includegraphics[width=1.7in]{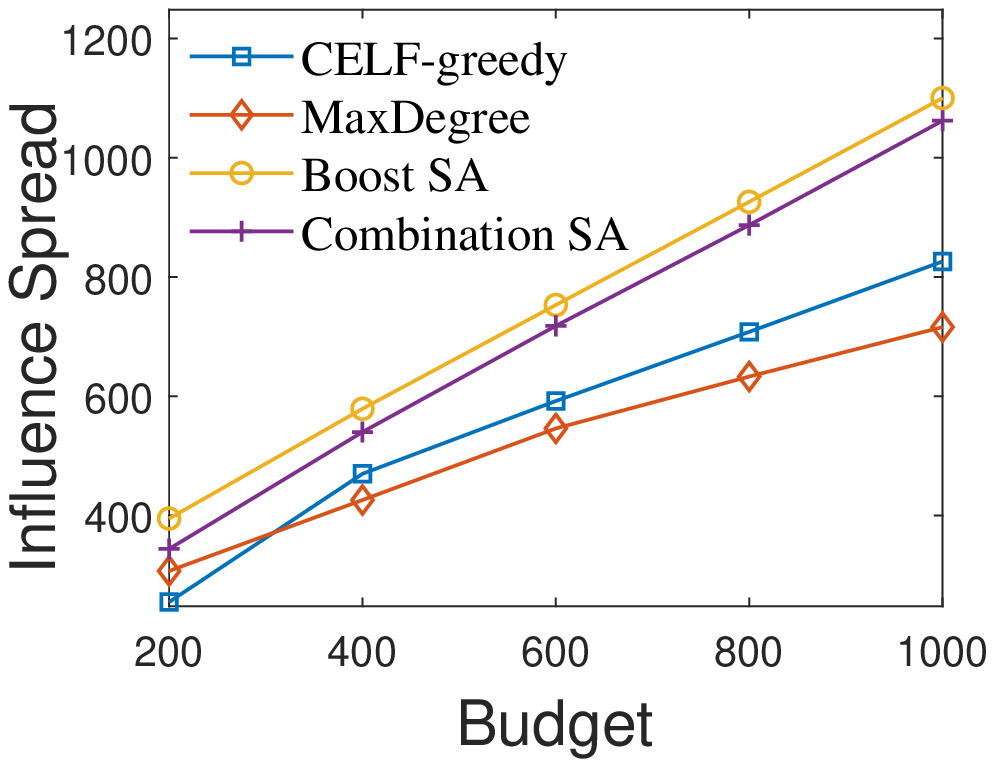}}
  \centerline{(a).network-2000}
\end{minipage}
\hfill
\begin{minipage}{0.24\textwidth}  
  \centerline{\includegraphics[width=1.7in]{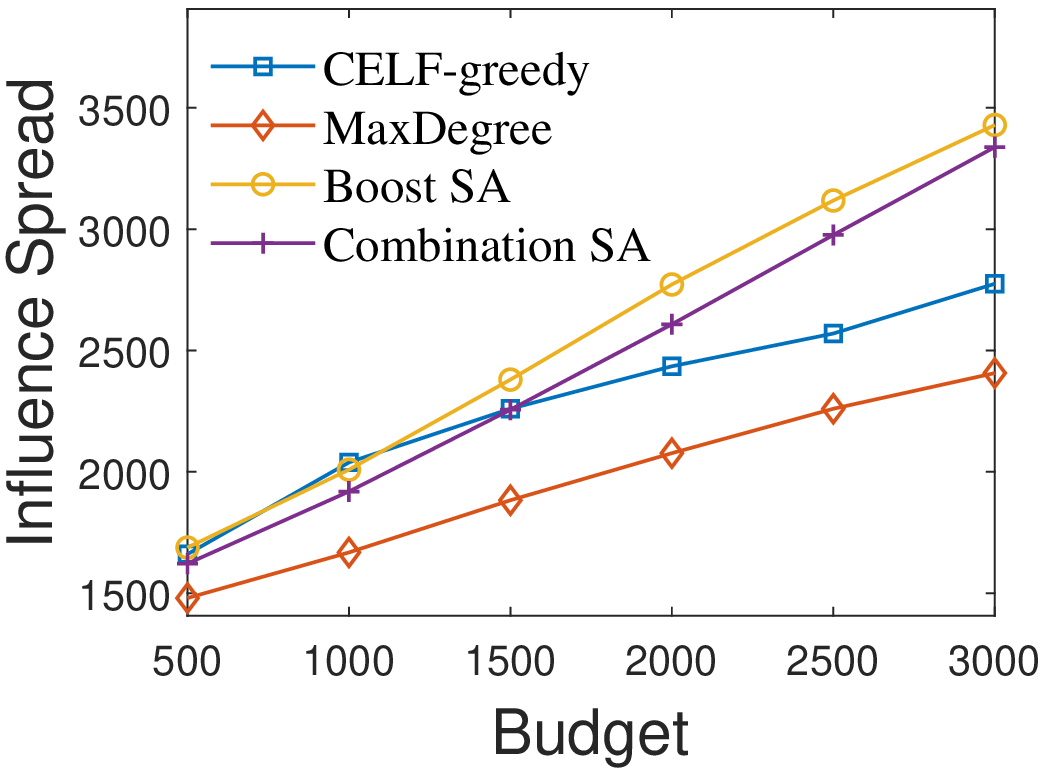}}
  \centerline{(b).network-5000}
\end{minipage}
\vfill
\begin{minipage}{0.24\textwidth}  
  \centerline{\includegraphics[width=1.7in]{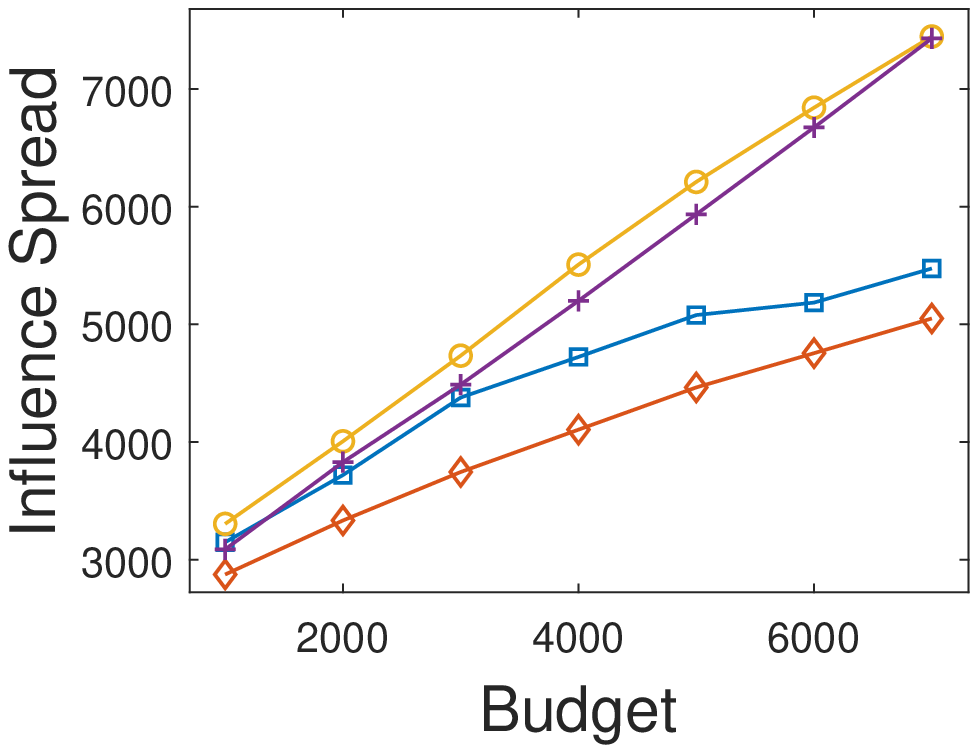}}
  \centerline{(c).network-10000}
\end{minipage}
\hfill
\begin{minipage}{0.24\textwidth}  
  \centerline{\includegraphics[width=1.7in]{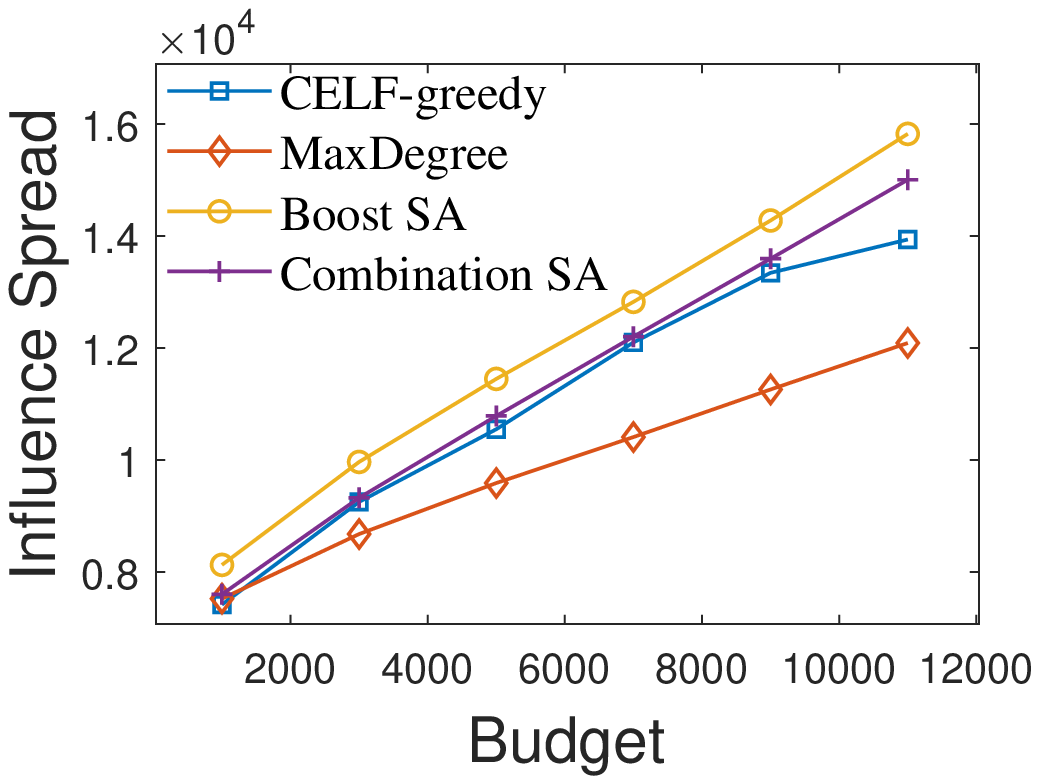}}
  \centerline{(d).network-30000}
\end{minipage}
\caption{Influence spread on  synthetic networks}
\end{figure}


\begin{figure}
	\centering
	\includegraphics[width=0.70\linewidth]{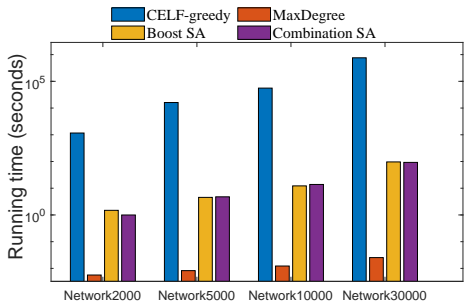}
	\caption{The runtime on synthetic networks}
	\label{fig8}
\end{figure}

\subsection{\textbf{Comparison between  the random initialization and unified initialization}}
In Algorithm \ref{alg:initial}, we randomly generate $k$ initial seed sets $S_i$ for generarte an initial seed set $S$ before sending it into the Boost SA. The random initialization has a larger search space than that of the uniform initialization.

To show the effects of the proposed random initialization in Algorithm \ref{alg:initial}, we keep the parameter settings described in subsection $C$, and compare the results of the Boost SA with the random initialization and unified initialization. The experimental results are shown in Tables \ref{table10}-\ref{table12} for three real world networks respectively. It can be seen  that the initialization strategy in Algorithm \ref{alg:initial} can promote the influence spread of the obtained seed set.

\linespread{1.5}
\begin{table}[!htbp]
	
	\centering
	\caption{Influence spread on URV email network}
	\label{table10}
	\begin{tabular}{|c|c|c|c|c|c|c|c|}
		\hline
		Budget&$100$&$200$&$300$&$400$&$500$&$600$&$700$\\
		\hline
		Random ini.& 425 & 490 & 563 & 656 & 740 & 782 & 842 \\
		\hline
		Unified ini.& 424 & 487 & 561 & 650 & 732 & 782 & 835 \\
		\hline
	\end{tabular}
\end{table}

\linespread{1.5}
\begin{table}[!htbp]
	
	\centering
	\caption{Influence spread on Wiki-Vote network}
	\label{table11}
	\begin{tabular}{|c|c|c|c|c|c|c|c|}
		\hline
		Budget&$1000$&$2000$&$3000$&$4000$&$5000$&$6000$ \\
		\hline
		Random ini.& 2169 & 2817 & 3651 & 4461 & 5506 & 6308 \\
		\hline
		Unified ini.& 2167 & 2801 & 3641 & 4458 & 5485 & 6305 \\
		\hline
	\end{tabular}
\end{table}

\linespread{1.5}
\begin{table}[!htbp]
	
	\centering
	\caption{Influence spread on NetHEPT network}
	\label{table12}
	\begin{tabular}{|c|c|c|c|c|c|c|c|}
		\hline
		Budget&$1000$&$2000$&$3000$&$4000$&$5000$&$6000$\\
		\hline
		Random ini.& 4435 & 5121 & 6005 & 6933 & 7636 & 10035 \\
		\hline
		Unified ini.& 4429 & 5119 & 6001 & 6924 & 7618 & 9965 \\
		\hline
	\end{tabular}
\end{table}


\subsection{\textbf{Nodes cost sensitivity analysis}}
\begin{figure}[H]
\begin{minipage}{0.5\textwidth}
  \centerline{\includegraphics[width=3.6in]{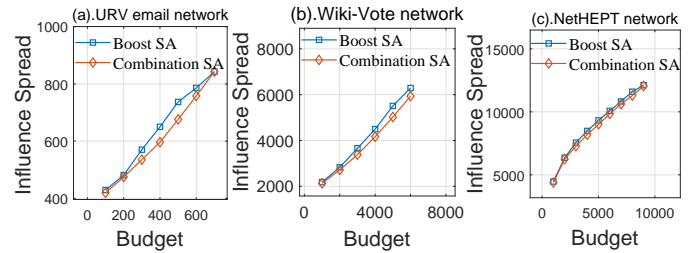}}
  \caption{Nodes cost sensitivity (98\% with $p=0.1$ and 2\% with $p=0.05$)}
  \label{fig:cost}
\end{minipage}
\end{figure}

The parameter $p$ is fixed as $p = 0.1$ in Eq.\ref{eq7}, which means that the activation cost of a node is positively correlated with its outdegree. In practical scenarios, this assumption is too absolute. In this subsection, we randomly select $2\%$ nodes and set  $p=0.05$, then compare Boost SA with Combination SA. The experimental results are shown in Fig.\ref{fig:cost}, which also shows that Boost SA always performs better than Combination SA. The results reflect that Boost SA can deal with more general network application problems.

\subsection{\textbf{Parameter sensitivity analysis}}
\begin{figure}[H]
	\begin{minipage}{0.5\textwidth}
		\centerline{\includegraphics[width=3.6in]{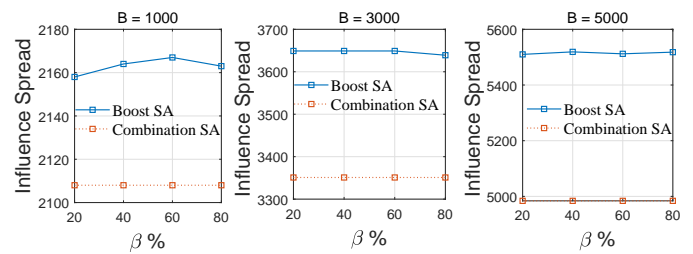}}
		\caption{Influence spread on Wiki-Vote network with different $\beta$ }
		\label{fig:beta}
	\end{minipage}
\end{figure}

\begin{figure}
	\centering
	\includegraphics[width=0.70\linewidth]{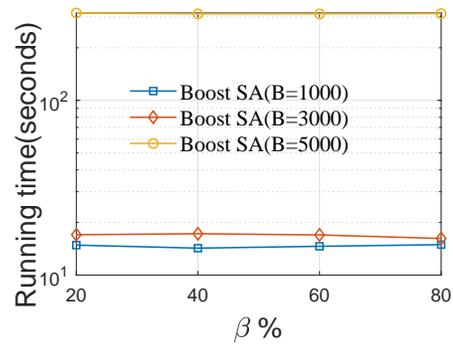}
	\caption{The runtime on Wiki-Vote network with different $\beta$ }
	\label{fig:beta_time}
\end{figure}

To investigate the effect of parameter $\beta$, we show the influence spread and running time results by varying $\beta$ on Wiki-Vote network in Fig.\ref{fig:beta} and Fig.\ref{fig:beta_time}. Generally speaking, nodes of $T_3$ in Fig.\ref{fig4} have more chances to be activated with the increasing of $\beta$. As shown in Fig.\ref{fig:beta}, in a wide range of $\beta$ from 20 to 80, Boost SA  works better than Combination SA, however, with larger value of $\beta$ the cost performance of the selected seed nodes may be falling. Fig.\ref{fig:beta_time} shows that the running time  is not sensitive with $\beta$.  Taking into account of different datasets and different budgets, we set $\beta = 60$.

\section{\textbf{Conclusion}}
Compared with the influence maximization (IM) problem, the budget influence maximization (BIM) is more approaching to the practical scenarios. In this paper, an efficient algorithm named Boost SA for the BIM problem is proposed, which is based on the simulated annealing algorithm.  Simulation results on real networks and synthetic networks show that the proposed algorithm can obtain a seed set with more influence spread with similar running time than state of the art algorithms. Without using a greedy strategy, the proposed Boost SA is scalable to large networks.\\

\bibliographystyle{IEEEtran}
\bibliography{IEEEabrv,bibfile}

\begin{thebibliography}{10}
\providecommand{\url}[1]{#1}
\csname url@samestyle\endcsname
\providecommand{\newblock}{\relax}
\providecommand{\bibinfo}[2]{#2}
\providecommand{\BIBentrySTDinterwordspacing}{\spaceskip=0pt\relax}
\providecommand{\BIBentryALTinterwordstretchfactor}{4}
\providecommand{\BIBentryALTinterwordspacing}{\spaceskip=\fontdimen2\font plus
\BIBentryALTinterwordstretchfactor\fontdimen3\font minus
  \fontdimen4\font\relax}
\providecommand{\BIBforeignlanguage}[2]{{%
\expandafter\ifx\csname l@#1\endcsname\relax
\typeout{** WARNING: IEEEtran.bst: No hyphenation pattern has been}%
\typeout{** loaded for the language `#1'. Using the pattern for}%
\typeout{** the default language instead.}%
\else
\language=\csname l@#1\endcsname
\fi
#2}}
\providecommand{\BIBdecl}{\relax}
\BIBdecl

\bibitem{Chen2010Scalable}
W.~Chen, C.~Wang, and Y.~Wang, ``Scalable influence maximization for prevalent
  viral marketing in large-scale social networks,'' in \emph{ACM SIGKDD
  International Conference on Knowledge Discovery and Data Mining}, 2010, pp.
  1029--1038.

\bibitem{Yu2015Friend}
Z.~Yu, C.~Wang, J.~Bu, X.~Wang, Y.~Wu, and C.~Chen, ``Friend recommendation
  with content spread enhancement in social networks,'' \emph{Information
  Sciences}, vol. 309, pp. 102--118, 2015.

\bibitem{Chen2020Efficient}
C.~Chen, M.~Zhang, W.~Ma, Y.~Liu, and S.~Ma, ``Efficient non-sampling
  factorization machines for optimal context-aware recommendation,'' in
  \emph{Proceedings of The Web Conference}, 2020.

\bibitem{2018In}
Y.~Li, J.~Fan, Y.~Wang, and K.~L. Tan, ``Influence maximization on social
  graphs: A survey,'' \emph{IEEE Transactions on Knowledge and Data
  Engineering}, vol.~30, pp. 1852--1872, 2018.

\bibitem{Feng2020Neig}
C.~Feng, L.~Fu, B.~Jiang, H.~Zhang, X.~Wang, F.~Tang, and G.~Chen,
  ``Neighborhood matters: Influence maximization in social networks with
  limited access,'' \emph{IEEE Transactions on Knowledge and Data Engineering},
  vol. early access, 2020.

\bibitem{Domingos2001Mining}
P.~Domingos and M.~Richardson, ``Mining the network value of customers,'' in
  \emph{ACM SIGKDD International Conference on Knowledge Discovery and Data
  Mining}, 2001, pp. 57--66.

\bibitem{Tardos2003Maximizing}
E.~Tardos, D.~Kempe, and J.~Kleinberg, ``Maximizing the spread of influence in
  a social network,'' in \emph{ACM SIGKDD International Conference on Knowledge
  Discovery and Data Mining}, 2003, pp. 137--146.

\bibitem{Hallier2013On}
H.~Nguyen and R.~Zheng, ``On budgeted influence maximization in social
  networks,'' \emph{IEEE Journal on Selected Areas in Communications}, vol.~31,
  no.~6, pp. 1084--1094, 2013.

\bibitem{Tong2020time}
G.~Tong, R.~Wang, Z.~Dong, and X.~Li, ``Time-constrained adaptive influence
  maximization,'' \emph{IEEE Transactions on Computational Social Systems},
  vol. early access, 2020.

\bibitem{Cai2020Tar}
T.~Cai, J.~Li, A.~Mian, R.~H.~A. Li, T.~Sellis, and J.~X. Yu, ``Target-aware
  holistic influence maximization in spatial social networks,'' \emph{IEEE
  Transactions on Computational Social Systems}, vol. early access, 2020.

\bibitem{2016Tar}
C.~Song, W.~Hsu, and M.~L. Lee, ``Targeted influence maximization in social
  networks,'' in \emph{Acm International on Conference on Information \&
  Knowledge Management}, 2016.

\bibitem{2017Most}
J.~Li, X.~Wang, K.~Deng, X.~Yang, and J.~X. Yu, ``Most influential community
  search over large social networks,'' in \emph{2017 IEEE 33rd International
  Conference on Data Engineering (ICDE)}, 2017.

\bibitem{2019Opinion}
Q.~He, X.~Wang, B.~Yi, F.~Mao, and M.~Huang, ``Opinion maximization through
  unknown influence power in social networks under weighted voter model,''
  \emph{IEEE Systems Journal}, vol.~PP, no.~99, pp. 1--12, 2019.

\bibitem{Leskovec2007Cost}
J.~Leskovec, A.~Krause, C.~Guestrin, C.~Faloutsos, J.~M. Vanbriesen, and N.~S.
  Glance, ``Cost-effective outbreak detection in networks,'' in \emph{ACM
  SIGKDD International Conference on Knowledge Discovery and Data Mining},
  2007, pp. 420--429.

\bibitem{Han2014Balanced}
S.~Han, F.~Zhuang, Q.~He, and Z.~Shi, \emph{Balanced Seed Selection for
  Budgeted Influence Maximization in Social Networks}.\hskip 1em plus 0.5em
  minus 0.4em\relax Springer International Publishing, 2014.

\bibitem{Chen2009Efficient}
W.~Chen, Y.~Wang, and S.~Yang, ``Efficient influence maximization in social
  networks. in: Kdd,'' in \emph{ACM SIGKDD International Conference on
  Knowledge Discovery and Data Mining}, 2009, pp. 199--208.

\bibitem{KUNDU2015107}
S.~Kundu and S.~K.Pal, ``Deprecation based greedy strategy for target set
  selection in large scale social networks,'' \emph{Information Sciences}, vol.
  316, pp. 107 -- 122, 2015.

\bibitem{Brin2012Reprint}
S.~Brin and L.~Page, ``Reprint of: The anatomy of a large-scale hypertextual
  web search engine,'' \emph{Computer Networks}, vol.~56, no.~18, pp.
  3825--3833, 2012.

\bibitem{SAITO2016985}
K.~Saito, M.~Kimura, K.~Ohara, and H.~Motoda, ``Super mediator �c a new
  centrality measure of node importance for information diffusion over social
  network,'' \emph{Information Sciences}, vol. 329, pp. 985 -- 1000, 2016.

\bibitem{Jiang2011Simulated}
Q.~Jiang, G.~Song, G.~Cong, Y.~Wang, W.~Si, and K.~Xie, ``Simulated annealing
  based influence maximization in social networks,'' in \emph{AAAI Conference
  on Artificial Intelligence}, 2011, pp. 127--132.

\bibitem{Metropolis2004Equation}
N.~Metropolis, A.~W. Rosenbluth, M.~N. Rosenbluth, A.~H. Teller, and E.~Teller,
  ``Equation of state calculations by fast computing machines,'' 1953.

\bibitem{Gong2016Influence}
M.~Gong, J.~Yan, B.~Shen, L.~Ma, and Q.~Cai, ``Influence maximization in social
  networks based on discrete particle swarm optimization,'' \emph{Information
  Sciences An International Journal}, vol. 367-368, no.~C, pp. 600--614, 2016.

\bibitem{Yang2018Influence}
J.~Yang and J.~Liu, ``Influence maximization-cost minimization in social
  networks based on a multiobjective discrete particle swarm optimization
  algorithm,'' \emph{IEEE Access}, vol.~6, no.~99, pp. 2320--2329, 2018.

\bibitem{Tang2015Influence}
Y.~Tang, Y.~Shi, and X.~Xiao, ``Influence maximization in near-linear time,''
  pp. 1539--1554, 2015.

\bibitem{Pan2020Multi}
G.~Panagopoulos, F.~D. Malliaros, and M.~Vazirgiannis, ``Multi-task learning
  for influence estimation and maximization,'' \emph{IEEE Transactions on
  Knowledge and Data Engineering}, vol. early access, 2020.

\bibitem{2020An}
X.~Deng, F.~Long, B.~Li, D.~Cao, and Y.~Pan, ``An influence model based on
  heterogeneous online social network for influence maximization,'' \emph{IEEE
  Transactions on Network Science and Engineering}, vol.~7, pp. 737--749, 2020.

\bibitem{Zhang2010Estimate}
Y.~Zhang, Q.~Gu, J.~Zheng, and D.~Chen, ``Estimate on expectation for influence
  maximization in social networks,'' in \emph{Pacific-Asia Conference on
  Advances in Knowledge Discovery and Data Mining}, 2010, pp. 99--106.

\bibitem{Zhu2014Maximizing}
T.~Zhu, B.~Wang, B.~Wu, and C.~Zhu, ``Maximizing the spread of influence
  ranking in social networks,'' \emph{Information Sciences}, vol. 278, pp.
  535--544, 2014.

\bibitem{Lee2014A}
J.~R. Lee and C.~W. Chung, ``A fast approximation for influence maximization in
  large social networks,'' pp. 1157--1162, 2014.

\bibitem{Hart2003The}
P.~Hart, ``The condensed nearest neighbor rule (corresp.),'' \emph{IEEE
  Transactions on Information Theory}, vol.~14, no.~3, pp. 515--516, 2003.

\bibitem{Ho1995Random}
T.~K. Ho, ``Random decision forests,'' in \emph{International Conference on
  Document Analysis and Recognition}, 1995, p. 278.

\bibitem{Guimer2003Self}
R.~Guimer��, L.~Danon, A.~D��az-Guilera, F.~Giralt, and A.~Arenas,
  ``Self-similar community structure in a network of human interactions,''
  \emph{Phys Rev E Stat Nonlinear \& Soft Matter Phys}, vol.~68, no. 6 Pt 2, p.
  065103, 2003.

\bibitem{Leskovec2010Signed}
J.~Leskovec, D.~Huttenlocher, and J.~Kleinberg, ``Signed networks in social
  media,'' in \emph{Sigchi Conference on Human Factors in Computing Systems},
  2010, pp. 1361--1370.

\bibitem{Andrea2009Benchmarks}
L.~Andrea and F.~Santo, ``Benchmarks for testing community detection algorithms
  on directed and weighted graphs with overlapping communities,''
  \emph{Physical Review E Statistical Nonlinear \& Soft Matter Physics},
  vol.~80, no.~2, p. 016118, 2009.

\end{thebibliography}

\end{document}